\documentclass[acmsmall,natbib=false]{acmart}

\usepackage{colortbl}

\usepackage{makecell}

\usepackage{amssymb}

\usepackage{colortbl}
\usepackage{amsmath}
\usepackage{graphicx}
\usepackage{xtab}
\usepackage{caption}
\usepackage{appendix}
\newtheorem{definition}{Definition}
\usepackage{url}
\AtBeginDocument{%
  }

\acmSubmissionID{CSUR-2023-0621}

\RequirePackage[
  datamodel=acmdatamodel,
  style=acmnumeric,
  ]{biblatex}

\begin{document}

\title{Methods for Acquiring and Incorporating Knowledge into Stock Price Prediction: A Survey}


\author{Liping Wang}
\email{lwang347@connect.hkust-gz.edu.cn}
\orcid{0009-0007-8673-2171}
\affiliation{%
  \institution{The Hong Kong University of Science and Technology (Guangzhou)}
 \city{Guangzhou}
 \state{Guangdong}
 \country{China}
}

\author{Jiawei Li}
\email{jli226@connect.hkust-gz.edu.cn}
\affiliation{%
 \institution{The Hong Kong University of Science and Technology (Guangzhou)}
 \city{Guangzhou}
 \state{Guangdong}
 \country{China}}

\author{Lifan Zhao}
\email{mogician233@sjtu.edu.cn}
\affiliation{%
  \institution{Shanghai Jiao Tong University}
  \city{Shanghai}
  \country{China}}

\author{Zhizhuo Kou}
\email{zkou690@connect.hkust-gz.edu.cn}
\orcid{}
\affiliation{%
 \institution{The Hong Kong University of Science and Technology (Guangzhou)}
 \city{Guangzhou}
 \state{Guangdong}
 \country{China}}

\author{Xiaohan Wang}
\email{1155178126@link.cuhk.edu.hk}
\orcid{}
\affiliation{%
 \institution{The Chinese University of Hong Kong}
 \city{Hong Kong}
 \state{Hong Kong}
 \country{China}}

\author{Xinyi Zhu}
\email{xzhu683@connect.hkust-gz.edu.cn}
\affiliation{%
 \institution{The Hong Kong University of Science and Technology (Guangzhou)}
 \city{Guangzhou}
 \state{Guangdong}
 \country{China}}

\author{Hao Wang}
\email{seraveea@connect.hkust-gz.edu.cn}
\orcid{0009-0008-6278-8709}
\affiliation{%
 \institution{The Hong Kong University of Science and Technology (Guangzhou)}
 \city{Guangzhou}
 \state{Guangdong}
 \country{China}}
 
\author{Yanyan Shen}
\email{shenyy@sjtu.edu.cn}
\affiliation{%
  \institution{Shanghai Jiao Tong University}
  \city{Shanghai}
  \country{China}}

\author{Lei Chen}
\email{leichen@cse.ust.hk}
\affiliation{%
\institution{The Hong Kong University of Science and Technology (Guangzhou)}
\city{Guangzhou}
\state{Guangdong}
\country{China}}

\renewcommand{\shortauthors}{Wang et al.}

\begin{abstract}
Predicting stock prices presents a challenging research problem due to the inherent volatility and non-linear nature of the stock market. In recent years, knowledge-enhanced stock price prediction methods have shown groundbreaking results by utilizing external knowledge to understand the stock market. Despite the importance of these methods, there is a scarcity of scholarly works that systematically synthesize previous studies from the perspective of external knowledge types. Specifically, the external knowledge can be modeled in different data structures, which we group into non-graph-based formats and graph-based formats: 1) \textit{non-graph-based knowledge} captures contextual information and multimedia descriptions specifically associated with an individual stock; 2) \textit{graph-based knowledge} captures interconnected and interdependent information in the stock market. This survey paper aims to provide a systematic and comprehensive description of methods for acquiring external knowledge from various unstructured data sources and then incorporating it into stock price prediction models. We also explore fusion methods for combining external knowledge with historical price features. Moreover, this paper includes a compilation of relevant datasets and delves into potential future research directions in this domain. 
\end{abstract}




\maketitle

\section{Introduction}
The field of stock prediction has grown significantly in recent years, with an increasing demand for more sophisticated and data-driven machine learning (ML) techniques that provide valuable insights. Traditional ML methods~\cite{hossain2018hybrid} predict stock market volatility relying on time-series data, which refers to historical prices and financial technical indicators. However, the high stochasticity and non-stationary properties of the stock price cause the data distribution to change over time~\cite{du2021adarnn,li2022ddg}. The technical indicators that we have learned in the past may well lose efficacy in the future. Additionally, the Efficient Market Hypothesis (EMH)~\cite{fama1970efficient} states that under ideal market conditions, stock prices reflect all information. However, in actual market conditions, it often takes time for valid market information to be fully transmitted to the stock price. Thus, by only utilizing time-series historical data for predictions is impossible to beat the market consistently. 

\begin{figure}[t]
    \centering
    \includegraphics[width=0.7\linewidth]{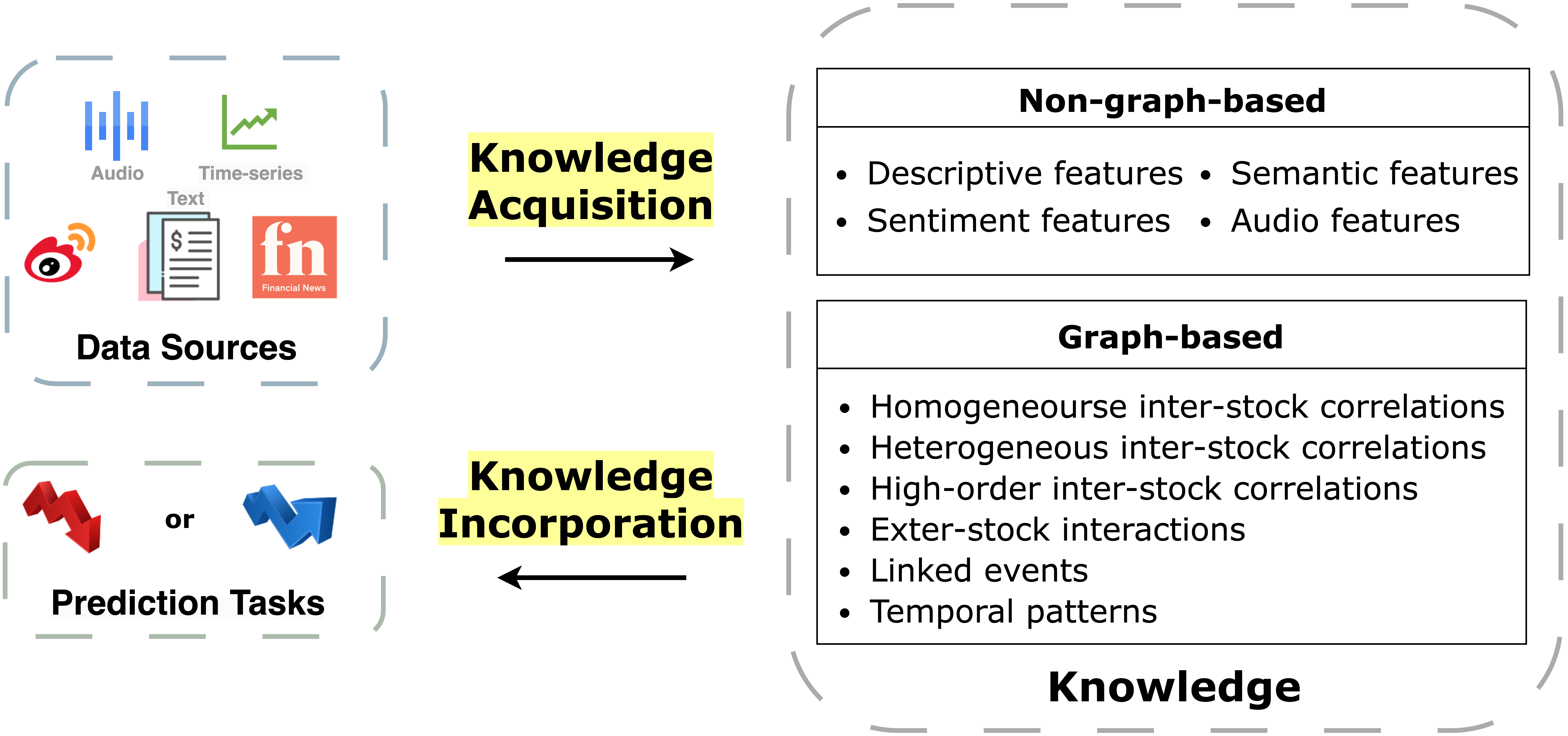}
    \caption{Formats of Knowledge and An Overview of Knowledge Acquisition and Knowledge Incorporation Operations.}
    \label{fig:1} 
    \vspace{-3mm}
\end{figure}

\textbf{Knowledge-enhanced methods} have emerged as a promising approach to overcoming the limitations of traditional stock prediction methods. This method leverages external knowledge, such as public sentiment, stock relationships, financial events, and intrinsic properties of a stock to understand its behavior. 
Traditional stock price prediction methods~\cite{akita2016deep} often overlook the interrelated nature of the stock market and treat each stock as an isolated entity. By incorporating the external knowledge, stock prediction methods are able to model the \textit{high-interconnected nature} of the stock market, which can be reflected in several key aspects: interdependence among listed companies, holding-relation from financial institutions to stocks, industry chain, spillover effects caused by events, etc. Knowledge-enhanced methods can also provide a more explainable solution that helps to understand the complex relations behind the stock price trends according to the background knowledge.
In this paper, the challenges of knowledge-enhanced stock price prediction methods are distributed in two aspects: knowledge acquisition and knowledge incorporation, which are essential to facilitate more accurate and reliable models (shown in Fig.~\ref{fig:1}).


\textit{Knowledge Acquisition}. Given the complexity of financial markets and the many factors that influence their dynamics, knowledge acquisition is necessary to enhance the quality of training data for prediction models. The external knowledge needs to be processed and organized to be easily integrated into the prediction models. 

\textit{Knowledge Incorporation}. The effectiveness of incorporation models lies in their ability to leverage signals that impact future prices from the knowledge. These models employ diverse techniques, such as deep learning models and graph-based models. In addition, the integration of historical time-series prices with external knowledge has been a prevalent approach in enhancing stock price prediction.


This survey paper focuses on providing an overview of the existing works associated with acquiring and incorporating external knowledge into stock price prediction models. 

\subsection{Differences with Existing Surveys}
Previous surveys mainly compare the technical differences among various stock price prediction models. Sonkiya et al.~\cite{sonkiya2022stock} divided the prediction models into five groups: Basic Linear Regressions Models, Traditional Machine Learning Models, Deep Learning Models, and Hybrid and Advanced Models. Hu et al.~\cite{hu2021survey} presented the recent literature on stock and foreign exchange (Forex) price prediction using deep learning methods, such as convolutional neural networks (CNNs), Long short-term memory (LSTM), deep neural networks (DNNs), recurrent neural networks (RNNs), Reinforcement Learning, and so on. Wang et al.~\cite{wang2021review} focused on GNN methods in financial applications, which applied graph data as alternative data to enhance financial applications. Saha et al.~\cite{saha2021stock} also delved into existing graph-based works for stock price prediction and portfolio optimization. Their discussion primarily focused on mathematical perspectives of constructing and preprocessing stock market graphs, including graph formulation, filtering, and clustering. 

In contrast, our survey is organized from the perspective of external knowledge. We categorize the external knowledge used in previous works into two distinct groups relying on data formats: non-graph-based knowledge group and graph-based knowledge group. Knowledge in the first group indicates contextual information and multimedia descriptions specifically associated with an individual stock. Knowledge belonging to the graph-based group can represent interconnected and interdependent information among multiple stocks in the stock market. Different acquiring and incorporating models are required to effectively leverage the unique characteristics of each knowledge type. This paper offers an extensive and in-depth description of methods for acquiring knowledge from various data sources and incorporating different types of knowledge into stock price prediction models. By thoroughly exploring these methods, we aim to equip researchers with comprehensive insights and practical guidance on effectively integrating external knowledge into their models. Moreover, previous surveys have failed to adequately emphasize the combination of time-series data and external knowledge. Recognizing this gap, this paper systematically classifies and analyzes fusion techniques.


\vspace{-1mm}

\subsection{Contributions and Structure of This Survey}

This survey encompasses the following contributions:
\begin{description}
\item[1.] We review the various types of knowledge that are utilized to enhance the performance of stock prediction models. We term them as \textit{non-graph-based knowledge} and \textit{graph-based knowledge}. (Sec.~\ref{sec:bac})
\item[2.] We review the models that \textbf{acquire knowledge} from various data sources, such as financial news and social media posts. We introduce the acquiring model in non-graph-based knowledge and graph-based knowledge separately. (Sec.~\ref{sec:acq})
\item[3.] We review the models that \textbf{incorporate knowledge} into stock prediction. In addition, based on the best of our knowledge, we categorize the techniques for combining time series and knowledge into two paradigms: sequential fusion and parallel fusion. (Sec.~\ref{sec:inc})
\item[4.] We provide a summary of the \textbf{existing financial-domain knowledge bases} and present a detailed statistical analysis of their characteristics. (Sec.~\ref{sec:data})
\item[5.] We provide research \textbf{challenges and future directions} for knowledge acquisition and knowledge incorporation, including AutoML for knowledge selection, addressing irregularity and granularity of financial time series, multimodal learning on stock knowledge, and knowledge-enhanced generalization against concept drift in the stock market. (Sec.~\ref{sec:fut})
\end{description}

\section{Background}\label{sec:bac}

\begin{figure}

\begin{minipage}{.52\linewidth}
\centering
\includegraphics[width=8cm]{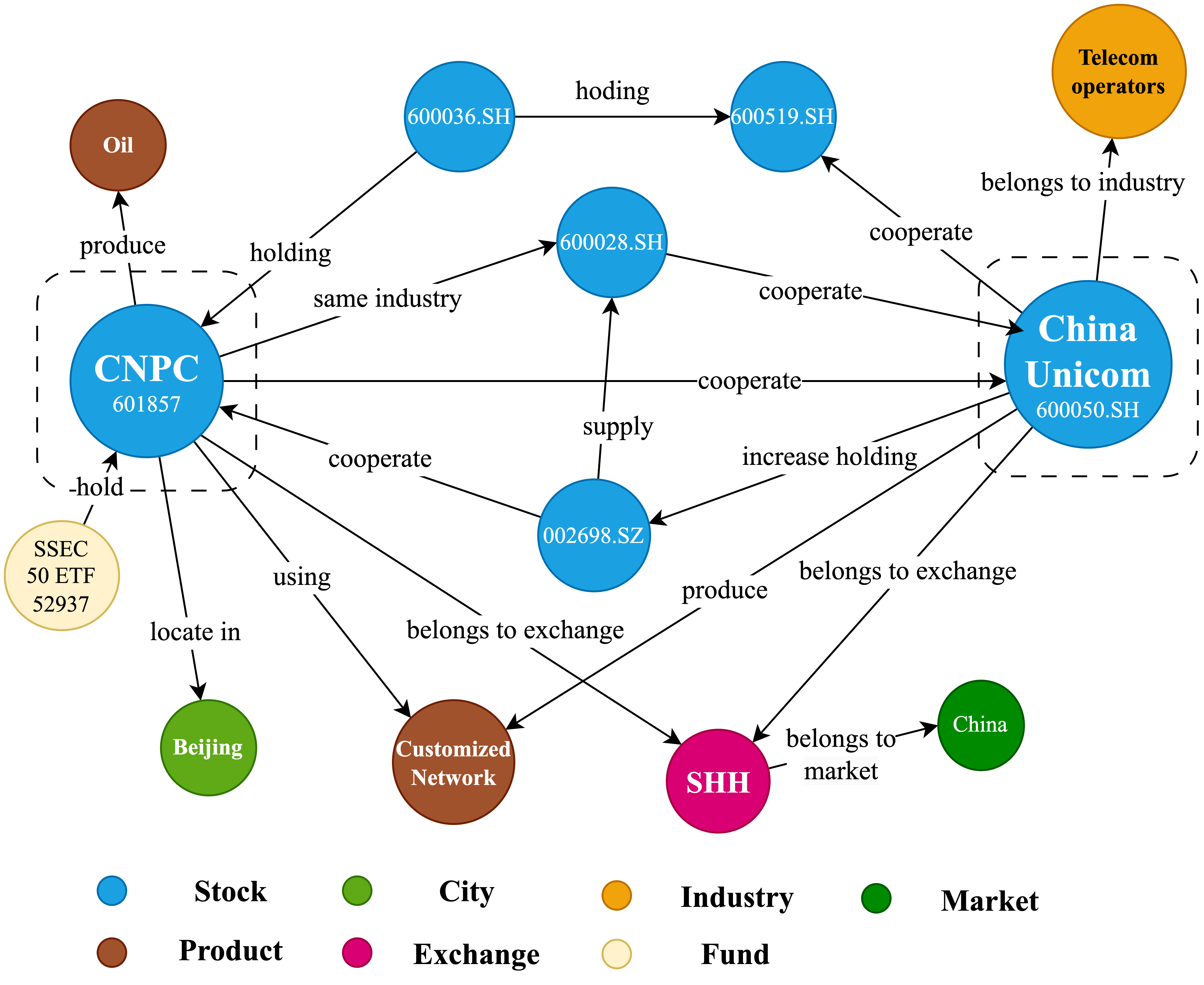}
\caption{An Example of Financial Graph from K-Quant~\cite{Kquant}}
\label{fig:2} 
\end{minipage}%
\begin{minipage}{.48\linewidth}
\centering
\captionof{table}{Notations Used in This Paper}\label{notation}
\resizebox{0.7\textwidth}{0.15\textheight}{
\begin{tabular}{cl}
\toprule
\bfseries Notations & \bfseries Descriptions \\
\midrule
 $P_i^t$ & Historical indicators\\&  of the $i$-th stock at\\&  the $t$-th day\\
 $event = (s,p,o)$ & Event tuple\\
 $v_i$ & Node / Entity $i$\\
 $e_{ij}$ & Edge / Link between \\& nodes $i$ and $j$ \\
$r_i$ & Relation $i$ \\
 $A$ & Adjacency matrix\\
 $\mathbf{X} = \{v_1,v_2,...,v_{|\mathcal{V}|}\}$ & A set of node features\\
 $\mathcal{V}$ & A set of nodes\\
 $\mathcal{E}$ & A set of edges \\
 $E$ & A hyperedge \\
$\mathcal{R}$ & A set of relation types \\
$\mathcal{S}$ & A set of node types \\ 
$\mathcal{G}$ & Static graph\\
$\mathbb{G}$ & Dynamic graph\\
$\tilde{\mathcal{G}}$ & Hypergraph\\

\bottomrule
\end{tabular}}
\end{minipage}
\end{figure}

This section first introduces the formulations of stock price prediction tasks and discusses why it is necessary to incorporate knowledge for stock price prediction. In addition, to prevent any confusion, the paper introduces the notations and concepts used in the article before delving into the advanced techniques. To make it easier to understand, Table~\ref{notation} contains a list of symbols and their corresponding explanations.

\subsection{Different Stock Price Prediction Tasks}

Stock price predicton problem can be treated as \textit{regression tasks}~\cite{xu2021rest,zhou2022temporal} or \textit{classification tasks}~\cite{deng2019knowledge,li2021modeling} or \textit{ranking tasks}~\cite{feng2019temporal,ying2020time}. In a regression task, the model predicts a numerical value for the stock price or one-day return ratio. In a classification task, the model outputs a binary label indicating the direction of the stock price movement: increase (1) or not increase (0). Graph-based knowledge-enhanced models can define this task as a node classification task~\cite{wang2022hatr}. Ranking tasks output the rankings of stocks based on their predicted scores.





\subsection{A Brief History of Stock Price Prediction} 

Many existing works in the past used temporal data gleaned from historical stock prices as fundamental data, such as open, high, low, and closed prices~\cite{nikou2019stock}. Apart from these indicators, several financial indicators were introduced by experienced investors and economists~\cite{zhu2009technical}. Researchers took into account these hand-crafted indicators in traditional stock price prediction~\cite{chung2018genetic} because these indicators are able to reflect the market trends to some extent. The hand-crafted indicators include the exchange rate, book-market ratio~\cite{chatzis2018forecasting}, trading volume~\cite{zhang2018new}, Exponential Moving Average (EMA), and Moving Average Convergence/Divergence (MACD)~\cite{li2019individualized}. Moreover, because macroeconomic statistics, such as CPI~\cite{tangjitprom2011macroeconomic} and GDP~\cite{ogundunmade2022stock}, reflect the economic growth of a certain region, several traditional works occasionally employed them in investment models. 

Prior to the emergence of machine learning methodologies, statistical techniques were commonly employed in traditional stock price prediction. These techniques included ARIMA~\cite{ariyo2014stock} and Exponential Smoothing Model (ESM)~\cite{de2009predicting}, along with their different variations. As machine learning technology advances, more and more scholars are concentrating on how machine learning helps stock price prediction. The traditional machine learning methods, such as SVM~\cite{fenghua2014stock}, decision trees~\cite{vaiz2016study}, and K-means~\cite{gupta2014clustering} perform well on stock prices and market forecasts. With the development of deep learning, CNNs, RNNs, and reinforcement learning techniques are widely used in stock price prediction~\cite{hossain2018hybrid,lu2020cnn,sim2019deep}. Currently, the knowledge-enhanced stock price prediction problem has become a thriving and potential topic, which also has witnessed groundbreaking developments.

\subsection{External Knowledge in Stock Price Prediction}
With advancements in information retrieval, recent works~\cite{kilimci2020efficient, liu2019transformer} have enabled the extraction of valuable knowledge from large volumes of unstructured text data related to the stock market. Financial alternative data sources include news articles, social media posts, financial reports, and audio data, which can be difficult and time-consuming to analyze manually. The valuable external knowledge includes public sentiment, stock relationships, financial events, and so on. The knowledge is stored in different data structures for efficiently incorporating into the prediction model. We divided them into non-graph-based formats and graph-based formats (shown in Fig.~\ref{fig:1}).

Non-graph-based knowledge encompasses information that is not structured in a graph format. Previous studies have utilized innovative sources of information, such as investors' sentiments~\cite{gupta2020sentiment} extracted from social media, as well as semantic features from financial news~\cite{sonkiya2021stock}. Additionally, researchers have investigated the impact of financial events on the corresponding stock prices~\cite{deng2019knowledge, xu2021rest}. However, within the realm of non-graph-based knowledge, the primary focus lies in textual content and multimedia descriptions specifically associated with an individual stock.

The introduction of the graph as a novel data structure to store knowledge has revolutionized the field of stock prediction models~\cite{bisarya2022stock,feng2019temporal,sarmah2022learning,wang2021review}. The graph structure stores interconnected and comprehensive financial knowledge, such as stock correlations, supply chains, and industrial interdependencies. For example, a relationship between two stocks can be represented as an edge in the graph in the form of \textit{(corporation A, relation, corporation B)}. Besides of above, the graph structure can also present implicit and potential knowledge. As illustrated in Fig.~\ref{fig:2}, the financial graph reveals that "CNPC" and "China Unicom" have numerous indirect connections. These connections imply that there is a potential for mutual influence or impact between the two companies. Despite being in different industries, the high degree of similarity between "CNPC" and "China Unicom" may mean that their stock prices are correlated, and changes in one may indicate a change in another. 

The development of graph-based incorporating methods has become increasingly crucial for understanding and analyzing the complex interconnected stock market~\cite{chen2018incorporating}. On the one hand, the utilization of graph-based knowledge enables stock prediction models to achieve higher prediction performance. On the other hand, graphs allow for the visualization of complex relationships and dependencies between entities in the stock market. By tracing the paths of decision-making within the graph, the factors and relationships that contribute to specific predictions can be uncovered. 

\section{Knowledge Acquisition}\label{sec:acq}

Owing to the inherent complexity and dynamic nature of the market, the acquisition of knowledge that accurately describes financial markets presents some challenges as follows. Firstly, compared to general knowledge acquisition, acquiring domain-specific knowledge requires additional efforts to discern and classify fine-grained boundaries between financial events. Secondly, the confidence and quality of the extracted financial knowledge are of paramount importance. Moreover, it is crucial to ensure that the extracted knowledge effectively captures and preserves the temporal and dynamic features that characterize the stock market. In this section, we are going to discuss how previous works acquire different financial knowledge from various data sources.

\subsection{Non-graph-based Knowledge}
The non-graph-based knowledge derived from various sources, including textual data (like news articles, social media posts, and industry reports) and audio data, has recently become an increasingly important source of alternative data for market analysis years~\cite{jiang2021applications}.
Financial news consists of reports from professional news outlets, official company announcements, and analyst recommendations and is often considered a reliable source of investment information. On the other hand, social media consists of user-generated content such as tweets, blog posts, and forum comments and is known for its real-time nature and high level of user engagement. Social media data is typically used to gauge public sentiment and interest in stock market-related issues~\cite{khan2020stock}. A variety of social media platforms might be useful for stock price prediction assignments. Weibo~\cite{chen2018leveraging} and Twitter~\cite{li2018stock} are frequently used by researchers to find remarks on particular stocks. Stock domain sites, like StockTwits\footnotemark[1]\footnotetext[1]{\url{https://stocktwits.com/}}, are also employed as financial text resources.

In order to utilize the knowledge in financial texts, the first step is to convert it into a format that can be read by machines. Text vectors can take the form of a vector representation of a single word or a document. It should be noted that a sentence, such as a news headline, can be considered a unique type of document. Conventional methods rely on word frequency counts to generate text vectors, referred to as descriptive features. With the advancement of machine learning, text vectors can now be learned from contextual information, resulting in more concise and abstract representations known as semantic features. 
Moreover, emotions and voice tones from the CEO talking in the earning conference meeting can also be considered for the stock price prediction task~\cite{yang2020html}. 

Based on the above, we define non-graph-based knowledge extracted from textual data and audio data as the following classes: \textbf{descriptive features}, \textbf{sentiment features}, \textbf{semantic features}, and \textbf{audio features}. In this section, we summarize solutions for acquiring the above types of knowledge separately.

\subsubsection{Descriptive features} Descriptive features are extracted based on pre-identified useful words or phrases in the textual data. The frequency of each word or phrase in an individual document is then calculated as its corresponding feature, and words that do not appear have features of zero. This yields high-dimensional and sparse text vectors at the document level, denoted as $v = [a_1, a_2, ..., a_n]$, where $a_i$ represents the frequency of the $i$-th word or phrase. To identify useful words or phrases, common methods include Bag-of-Words (BoW)~\cite{groth2011intraday}, N-Gram~\cite{butler2009financial}, N-word combination~\cite{hagenau2013automated}, etc. 

\subsubsection{Sentiment features}

Sentiment features are obtained by sentiment analysis of text data to reflect investors' attitudes toward the market. Data sources suitable for acquiring sentiment feature is social media including Twitter, news websites, Sina Weibo, StockTwits, etc. In line with Gupta et al.~\cite{gupta2020sentiment}, research on financial public sentiment analysis can be categorized into lexicon-based or machine learning-based methods.

For lexicon-based methods, the researcher needs to use a dictionary related to public sentiment to convert each sentiment keyword appearing in the text into a scalar or vector. Li et al.~\cite{li2018stock} utilized the Loughran-McDonald financial sentiment dictionary~\cite{loughran2011liability} to track the occurrence of relevant sentiment keywords in news texts, converting each word into a 6-dimensional sentiment vector. Each dimension corresponds to a polarity in different aspects of sentiment. To better adapt to the financial texts, Wang et al.~\cite{wang2018combining} expanded the Sentiment Dictionary by adding financial terms and financial adjectives. The sentiment feature of text $i$ can be obtained by multiplying the frequency of keyword $j$ with its corresponding sentiment vector. 



\begin{equation}
\begin{aligned}
\textit{Sentiment Feature } (i) = \sum_{j\in\text{Lexicon}} Freq_j\cdot Vector_j
\end{aligned}
\end{equation}

However, lexicon-based techniques lack the support to deal with new words effectively because of the static sentiment dictionary~\cite{gupta2020sentiment}. In machine learning-based methods, the researchers commonly treat the sentiment analysis problem as a binary classification task, categorizing the text into positive or negative. Several works~\cite{chen2018leveraging, gupta2020sentiment} utilized Logistic Regression and SVM to predict the label of each tweet. Daily sentiment values are then calculated based on the difference between the number of positive tweets and the number of negative tweets. 
Moreover, in order to capture the volatility of the market, existing research constructs a sequence of sentiment scores on a daily basis corresponding to the daily stock prices. However, for many stocks on some trading days, there are possible to have more than one piece of news or non of one news. Particularly for less popular stocks, tweets or news related to certain stocks may not be available on some days. This causes the problem of irregular time series in sentiment scores, such as missing values and unbalanced sentiment distribution, which presents challenges. In such cases, Wang et al.~\cite{wang2018combining} employed a regression forecasting method (e.g., ARIMA) to generate the missing value. Li et al.~\cite{li2018stock} shared the sentiment feature space based on the similarity between stocks so that the news of popular stocks can also have an impact on unpopular stocks.

\subsubsection{Semantic features}
Researchers have leveraged this valuable knowledge to effectively capture the contextual and semantic information embedded within textual data, encompassing diverse sources like news content, news titles, and social media. By employing natural language processing (NLP) techniques, they adeptly transform the text into fixed-length vectors, enabling numerical computations. Two prominent categories of methods have emerged in this domain: machine learning methods (e.g., Word2vec~\cite{kilimci2020efficient, lee2017predict} and GloVe~\cite{kilimci2020efficient}) and deep learning methods (e.g.,  Transformer~\cite{chen2022gated}, BERT~\cite{chen2021stock}, FinBERT~\cite{sonkiya2021stock}). However, a particular financial term may trigger different effects on related stocks depending on the specific context. Thus the static embedding cannot well represent the text. For example, the word "acquisition" may have a positive connotation when the acquisition is beneficial to the development of the company or a negative connotation when the acquisition price is unreasonable. Several works dedicated to tackling this problem. For example, Huang et al.~\cite{huang2022news} adopted ELMo, a bi-directional LSTM-based model to learn the non-static representation of financial terms. 

\subsubsection{Audio features}

Audio features can be obtained from audio recordings and corresponding transcripts~\cite{medya2022exploratory}. The earnings call of publicly traded companies, widely regarded as a major event, is associated with a surge in stock market volatility and trading volume before, during, and after the event~\cite{medya2022exploratory}. The CEO or management representatives typically report financial achievements from the last quarter and provide guidance for the future during earnings calls~\cite{yang2020html}. However, audio data is inherently noisy, posing difficulties for the knowledge acquisition process. Yang et al.~\cite{yang2020html} discarded more than half of the data to ensure the quality of the remaining data, while Qin et al.~\cite{qin2019you} selected only the sentence made by the most spoken executive (usually the CEO) to avoid interference among different speakers. They used the pre-trained model to obtain the word embedding of the transcripts and used Praat~\cite{boersma2001speak} to obtain vocal features, such as pitch, intensity, and vocal tone. After that, they made the word embeddings and vocal features aligned at the sentence level.

\subsection{Graph-based Knowledge}

\begin{figure}
\begin{minipage}{.5\linewidth}
\centering
\includegraphics[width=7.2cm]{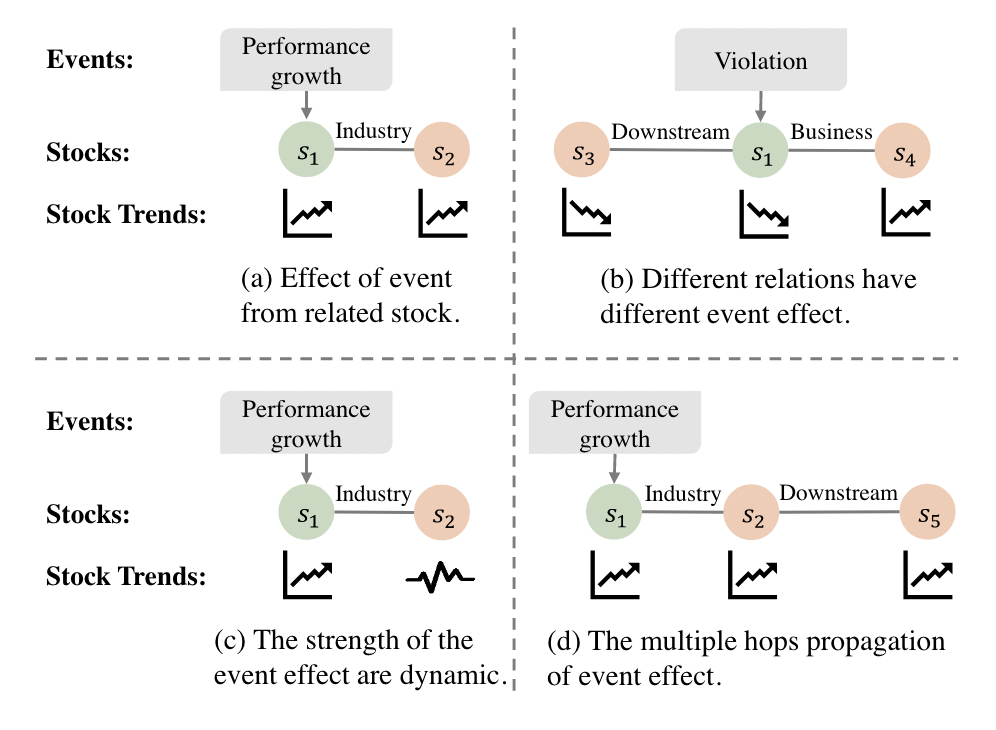}
\caption{An Observation in the Potential Impact of Events and Different Relationships on Trends of Correlated Stocks~\cite{xu2021rest}} \label{fig:event_effect} 
\vspace{-1mm}
\end{minipage}%
\begin{minipage}{.45\linewidth}
\captionof{table}{Types of Knowledge Contained in the \\Heterogeneous Inter-stock Correlation Network}\label{multi-acq}
\centering
\footnotesize
\begin{tabular}{cl}
\toprule
 \bfseries Ref. & \bfseries Relations   \\
\midrule
 ~\cite{xiang2022temporal} &  Positive \&  Negative Correlations\\
 ~\cite{cheng2020knowledge} & Investment \& Employment,\\&  Downstream \& Upstream, \\ &  Cooperation \& Competition \\
 ~\cite{xu2021rest}& Industry, Business, Shareholder, \\ & Downstream \& Upstream \\
 ~\cite{feng2019temporal} \& \cite{matsunaga2019exploring} & Industry,  \\ & Wiki Company-based Relations  \\
\cite{zhao2022stock} & Industry, Supply Chain, \\ & Partnership,  Investment\\
\bottomrule
\end{tabular}
\vspace{-1mm}
\end{minipage}
\vspace{-3mm}
\end{figure}

A large amount of financial knowledge can be effectively stored and represented using a graph structure. Each company is represented as a node in the graph, and the links define the companies' relationships. For example, the upstream and downstream knowledge can be represented by the corresponding relationships between companies such as (\textit{$company\_A$}, $is\_supplier$, \textit{$company\_B$}). And the shareholder knowledge of company A can be represented by (\textit{$company\_A$}, $is\_shareholder$, \textit{$company\_B$}). 
The stock-related knowledge encapsulated within a knowledge graph can be bifurcated into two primary categories: internal or external knowledge. \textbf{Internal stock knowledge} refers to correlations among stocks, such as cooperation, competition, ownership, upstream and downstream connections between stocks, etc. \textbf{External stock knowledge} refers to the relations between stock and other entity sets, including stock-industry relationships, industry chain connections, company location, exchange affiliations, etc. 



According to Fig.~\ref{fig:event_effect}, different relationships can have varying effects on market trends and stock prices, and an event has the potential to generate a commensurate or even more substantial effect on the stocks that lack an explicit correlation. For example, considering the correlation among stocks (inter-stock correlation), a competitive relationship between two companies can lead to negative spillover effects, where an increase in the stock price of one company results in a decrease in the stock price of the other due to competition for market share, resources, and customers. 
Moreover, relations between stock and external entities (exter-stock relations) may lead to a different effect on stock prices, such as stock-industry relations. If there is a positive industry trend, even companies in competition with each other in the same industry may experience an increase in their stock prices at the same time. Meanwhile, inter-stock correlation can be homogeneous or heterogeneous according to the number of relation types considered between stocks. The stocks that have the same attributes can be grouped into a collective group. Group-wise relationships can model high-order interdependence knowledge among stocks. Additionally, in real-world applications, the volatility of the financial market is a significant feature since corporation relations can be continuously evolving. 

\begin{figure}[!t]
    \centering
    \includegraphics[width=0.8\linewidth]{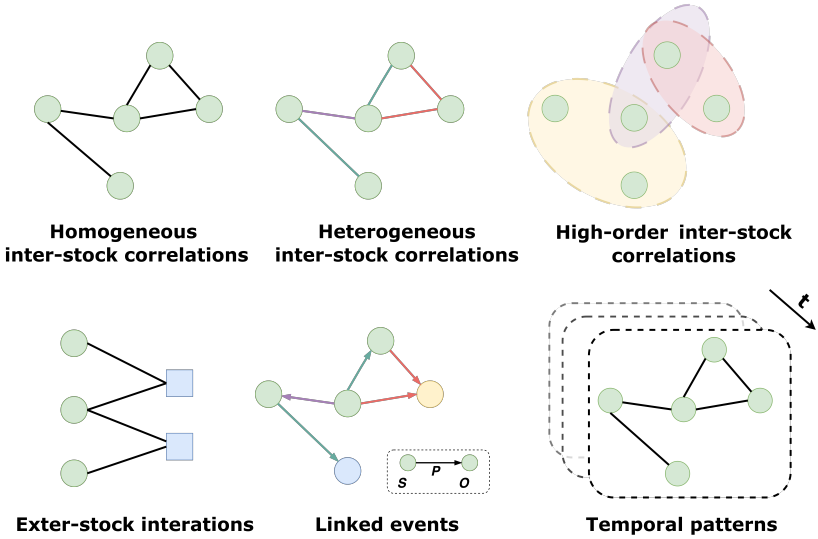}
    \caption{Demonstration of Each Graph-based Knowledge. (Nodes represented as circles denote stocks, while cubes represent concepts that interact with those stocks. The links are colored to indicate different types of relations.)
    }
    \label{fig:know} 
    \vspace{-5mm}
\end{figure}

Based on the above facts, various types of knowledge can be leveraged to improve stock market movement prediction. Thus, researchers have explored graph structures to aggregate different kinds of knowledge in one structured data for comprehensively representing the stock network. As shown in Fig.~\ref{fig:know}, the stock-related knowledge can be categorized into five types of graph-based knowledge: \textbf{homogeneous inter-stock correlations}, \textbf{heterogeneous inter-stock correlations}, \textbf{high-order inter-stock correlations}, \textbf{exter-stock interactions}, \textbf{linked events}, and \textbf{temporal patterns}.



 

\subsubsection{Homogeneous inter-stock correlations}\label{sec:homoac}

Homogeneous inter-stock correlation knowledge is related to similarities and interdependencies among stocks, which can be modeled by the homogeneous graph structure. In a homogeneous graph, nodes and edges are of the same type, which constructs a stock market network in the simplest graph format. Thus, the stock correlation knowledge can be modeled as follows. 

\begin{definition}{Homogeneous stock correlation network.} $\mathcal{G} = (\mathcal{V},\mathcal{E})$, where $\mathcal{V}$ is a set of nodes representing stocks. If two stock nodes ${v_i,v_j}$ have a relation, they are linked by an edge $e_{i,j}\in \mathcal{E}$. 
\end{definition}




The construction methods are different according to the implicit or explicit knowledge considered. 
\textbf{Implicit} homogeneous inter-stock correlation knowledge is generally modeled based on the \textit{correlation matrix} or \textit{similarity matrix}. Li et al.~\cite{li2021modeling} built a stock correlation graph where the nodes represent stock. The nodes are interconnected based on a correlation matrix that is computed using the historical market price. There will be an edge between node $i$ and $j$, if they have a high correlation coefficient. 
It is prevalent to use Pearson’s correlation coefficient and its derivatives to formulate the correlation of stock pairs based on returns~\cite{raffinot2017hierarchical,sarmah2022learning} and close prices~\cite{chi2010network} of stock pairs. Besides Pearson’s correlation measurement, the conditional Granger causality index (CGCI) is used to determine the causal relationship between two stocks. Mutual information (MI) and its derivatives~\cite{yan2020development} are used as similarity measures between stocks to construct the homogeneous stock graph. Both unweighted and weighted stock graphs can be constructed by the correlation matrix. For the unweighted stock correlation graph, existing works~\cite{li2021modeling} set a threshold to determine the positively correlated and negatively correlated relationship between nodes to reduce the noise of the correlation matrix. There will be an edge between node $i$ and $j$, if the absolute value of the correlation score is above a threshold $|\rho_{ij}| \ge \rho_{threshold}$.
There are various methods for constructing a weighted graph based on the correlation matrix. The weight can be set by the absolute value of the correlation coefficient~\cite{sarmah2022learning}, exponentially weighted correlation coefficient~\cite{pozzi2013spread}, and the largest eigenvalue of the correlation matrix~\cite{li2019portfolio}. To prevent the over-dense connection in the weighted fully-connected graph, algorithms like minimum spanning tree (MST)~\cite{aste2010correlation} and planar maximally filtered graph (PMFG)~\cite{pozzi2013spread} are used to remove "unimportant" links. 

\textbf{Explicit} homogeneous inter-stock correlation knowledge is formulated according to the \textit{explicit information} of listed companies, such as industry and main business areas information. For example, if a stock entity pair comes from the same industry, then there will have an edge between them. Sawhney et al.~\cite{sawhney2020deep} constructed a stock relation graph using the relation of companies stated by Wikidata\footnotemark[2]\footnotetext[2]{\url{https://www.wikidata.org}}.
They extracted two hop relations among stocks in the S\&P 500 index from Wikidata. If a first or second-order relation links two stocks, $v_i$ and $v_j$, there will be an edge $e_k = (v_i,v_j)$ between them. TRAN~\cite{ying2020time} obtained stock correlation knowledge based on explicit industry information. They converted an official industry classification list of stocks into a multi-hot binary matrix $C$. If $c_i \cdot c_j \geq  1$, a connection is established between stock $i$ and $j$.

%


\subsubsection{Heterogeneous inter-stock correlations}


Heterogeneous inter-stock correlations represent diverse relations of the individual stock, which can be modeled by the multi-relational graph structure. The multi-relational graph is a graph with different types of relations, where each edge connects exactly two entities. The heterogeneous inter-stock relation network can be defined as:
\begin{definition}{Heterogeneous inter-stock correlation network.} $\mathcal{G} = (\mathcal{V},\mathcal{E})$, where $\mathcal{V}$ is a set of nodes representing stocks. The edges $\mathcal{E}$ are extended by a set of relation types $\mathcal{R}$ to $ ( e, r ) \ \forall \ e \in \mathcal{E}$ of edge type $r \in \mathcal{R}$. Each edge in the graph can be formed as $<v_i,r,v_j>$. The number of relation types $|\mathcal{R}|$ is larger than 1.
\end{definition}

In contrast to the homogeneous knowledge mentioned previously, acquiring heterogeneous inter-stock knowledge aims to represent various relationships between stocks while preserving their contextual meaning. The relationships between companies can be complex and may even conflict with one another. For example, companies have competitive or cooperative relationships that may lead to differing effects on the target stock. We provide some frequently used relation types in Table \ref{multi-acq}. 

To acquire heterogeneous inter-stock knowledge, using explicit information is indeed a common method, as described in the stock correlation section (Sec.\ref{sec:homoac}). While explicit information can be a valuable source of data, it may not allow for the self-defined relation types in a heterogeneous relation network. To address this challenge, some researchers have employed relation extraction models to extract the desired relations from multiple sources of text data. These models leverage techniques such as NLP, named entity recognition, and machine learning algorithms. Cheng et al.~\cite{cheng2020knowledge} presented a novel approach for stock price prediction using a knowledge graph-based event embedding framework (KGEEF). The authors argued that an event could have effects not only on the primary company involved but also on related companies, which can be attributed to the lead-lag relationship. They stated that this lead-lag relationship is more prominent among manufacturing enterprises. Thus, KGEEF leveraged a relation detection model to construct inter-relations of stocks including upstream and downstream relations in the industry chain, investment/employment, and cooperative/competitive relationships. Based on the extracted inter-stock correlation, the propagation path of the event impact is determined.

On the other hand, Xiang et al.~\cite{xiang2022temporal} argued that extracting financial knowledge directly from stock prices can be more efficient and less noisy than acquiring it from textual sources. In light of this, the authors developed a multi-relation graph based on the correlation matrix, where the relations represent positive and negative correlations between stocks. Companies may have either a positive relation (if their correlation exceeds a predetermined threshold) or a negative relation (if their correlation falls below the threshold). In order to reduce noise, edges are established only between companies whose absolute correlation value surpasses the threshold, while the remaining edges are not established as connections.

There are several challenges revealed in acquiring heterogeneous inter-stock correlations. The first challenge is that the effectiveness of relation extraction models is contingent on the quality and quantity of training data. Additionally, the models may have difficulty detecting nuanced or subtle relationships between entities. Therefore, it is crucial to select appropriate sources of text data and fine-tune the model to the specific task at hand. To address this challenge, Zhu et al.~\cite{zhu2023hit} improved the distantly supervised relation extraction module by putting human-interact information in the loop.

\subsubsection{High-order inter-stock correlations}
High-order inter-stock correlations can represent group-wise relations between stocks, which can be molded by the hypergraph. As a generalization of graphs, hypergraphs can effectively represent collective higher-order relations between multiple stocks (or companies) simultaneously through hyperedges. Unlike traditional graphs, which only allow pairwise relationships between entities, hypergraphs can capture relationships between any number of entities through hyperedges. By utilizing hypergraphs, researchers can better model the complex correlations and dependencies among the stock groups and individual stocks. 
\begin{definition}{High-order inter-stock correlation network.}
    $\tilde{\mathcal{G}}=({\mathcal{V}}, {\mathcal{E}})$, where $\mathcal{V}$ is a set of stock nodes. Each hyperedge $E_i \in \mathcal{E} = \{v_1,v_2,..v_n\}$ represents a subset of related stocks.  
\end{definition}

Acquiring high-order inter-stock correlation knowledge requires a strong foundation in domain knowledge. Previous studies have used high-confidence explicit information to categorize stocks into different group sets. The industry sector is a natural way to classify stocks since companies within the same sector are affected by similar government policies, laws, and tax rates. Huynh et al.~\cite{huynh2023efficient} and Li et al.~\cite{Li2022hyper} constructed a hypergraph that each hyperedge connects the stocks in the same industry. Except for the industry information, some works utilize well-known open-domain datasets and selected stock-related information. Sawhney et al.~\cite{sawhney2021stock} used two data sources: industry and Wiki company-based data. They constructed three kinds of hyperedges, named same industry, first-order relations, and second-order relations. The first-order relations and second-order relations hyperedges represent the directly connected stocks and the stocks connected with the same entity respectively. 

Furthermore, relying solely on predefined concepts, such as industry classification, which are determined by human experts, may result in incomplete representations of the complex interdependencies among stocks. Emerging and critical concepts may not be promptly incorporated, leading to potential gaps in the analysis~\cite{huynh2023efficient}. Huynh et al.~\cite{huynh2023efficient} addressed this issue by considering similarity measures between stocks based on their historical stock prices from the previous year. They used clustering techniques to simulate market lag, where a leading stock influences the trend of other stocks. They constructed hyperedges that consist of stocks within the same cluster. To reflect the significance of each hyperedge, they derived the weight of each hyperedge from the total market capitalization of all related stocks.

\subsubsection{Exter-stock interactions}
Exter-stock interactions knowledge refers to interactions between two disjoint node sets, such as the holding relationship between funds and stocks, and the affiliation between stocks and industries. The bipartite graph structure can represent these interactions naturally. In a bipartite graph, nodes are divided into two independent groups and no connection between nodes in the same group. The exter-stock interaction knowledge can be formed as: 
\begin{definition}{Exter-stock interaction network.} For the graph $\mathcal{G} = (\mathcal{V},\mathcal{E})$, the node set $\mathcal{V}$ can be further partitioned into two disjoint subsets $\mathcal{V}_A$ and $\mathcal{V}_B$, where $\mathcal{V}_A\cup \mathcal{V}_B = \mathcal{V}$ and $\mathcal{V}_A\cap \mathcal{V}_B = \emptyset$. Every edge $e\in \mathcal{E}$ connects a node in $\mathcal{V}_A$ to a node in $\mathcal{V}_B$. $\mathcal{V}_A$ represents listed companies.
\end{definition}

There are three types of exter-stock interaction knowledge that are wildly used in financial analysis: investor-stock bipartite knowledge, concept-stock bipartite knowledge, and employee-stock bipartite knowledge. 

\textbf{Investor-Stock interaction knowledge} represents the interaction between stock and investors, such as funds~\cite{li2019individualized,zhou2022temporal}, fund institutions~\cite{lu2018herding}, and other institutional investors~\cite{bisarya2022stock}. The assumption for establishing a fund-related graph is that stocks within the same fund are likely to share some common characteristics~\cite{li2019individualized}. When an investor holds a stock for a certain period of time, there will be an edge connecting the investor and the stock. The edge weight in the bipartite graph refers to assigning a value or ratio to each edge to reflect the investment characteristic and can represent the strength of connections between nodes. For example, values of weights can be determined by the amount of stock market value owned by investor~\cite{lu2018herding}, the distribution that the fund invests on the stock~\cite{li2019individualized}, and shareholding ratio~\cite{bisarya2022stock}. The shareholding ratio is denoted as the ratio between the shares held by an institution to the total outstanding shares of the company. Lu et al.~\cite{lu2018herding} obtained explicit investment data for the fund from an open-source platform. Based on the assumption of collective actions among mutual funds under the same management institution, they categorized funds into groups by mutual fund management institutions. Instead of establishing multiple links between funds and their respective holdings, they established links solely between the management institutions and the stocks. An edge between institution $m$ and stock $i$ exists if the institution holds the stock, with weight $w_{mi}$ representing the market value of stock $i$ owned by institution $m$.


\textbf{Concept-Stock interaction knowledge} represents the interactions between stock and predefined stock concepts, such as sector, industry, and main business area. When two companies are in the same industry, their stock values may follow similar patterns and tend to be affected by similar market shocks. Feng et al.~\cite{feng2019temporal} collected 112 and 130 types of industry for NASDAQ and NYSE stocks from the official company list\footnotemark[3]\footnotetext[3]{\url{https://www.nasdaq.com/screening/industries.aspx}}. They connected stock pairs within the same industry. However, the industry relationship data is sparse since less than 10\% of stock pairs have at least one type of this relation~\cite{feng2019temporal}. Xu et al.~\cite{xu2021hist} added the main business data of companies (e.g., cloud computing, e-commerce) obtained from Tushare\footnotemark[4]\footnotetext[4]{\url{https://tushare.pro/.}}.


\textbf{Employee-Stock interaction knowledge} represents interactions between stock and people, such as board members, and managers. Investor-company or concept-company relations can reflect the external environment of the company, while person-company relations can reflect the internal corporate governance structure and can be used to analyze relationships between individuals and organizations. Creamer et al.~\cite{creamer2009link} introduced directors and analysts to a person-stock bipartite graph and predicted the earnings using node statistics. Additionally, executive relationships such as those formed by alumni networks have an implicit impact on the future decision of the listed companies. Zhao et al.~\cite{zhao2022stock} have additionally integrated the connections among executives in different companies and inter-stock relations to enrich the employee-stock interactions.

\subsubsection{Linked events}
Linked events knowledge can represent an overview of the stock market by connecting event tuples that have the same entity. Event tuple, denoted by $event = (s, p, o)$, comprises the action or predicate $p$, the actor or subject $s$, and the object $o$ on which the action is performed. Events are extracted from the textual data, generally financial news, and then structurize to the tuple structure. The linked event knowledge can be modeled by the heterogeneous graph, which contains various types of edges. The type of entities can also be heterogeneous since the types of subjects and objects can be different. The linked event knowledge can be modeled as follows:



\begin{definition}{Linked event network.}
For the graph $\mathcal{G} = (\mathcal{V},\mathcal{E})$, the node set $\mathcal{V}$ is determined by $\mathcal{V} \subseteq \mathbb{N} \times \mathcal{S}$ with a node type set $\mathcal{S}$. Thus, a node $(v,s)\in \mathcal{V}$ is defined by itself $v$ and its types $s$. The edge is extended by a set of relation types $\mathcal{R}$. The number of relation types $|\mathcal{R}| > 1$ and the number of node types $|\mathcal{S}| \geq 1$.
\end{definition}

The limited development in the field of heterogeneous graph work may have resulted in fewer attempts to incorporate this aspect into financial analysis models. Previous research~\cite{deng2019knowledge,cheng2020knowledge} indicates that the linked event knowledge acquisition can be split into two stages: \textit{event extraction} followed by \textit{structuralization}. Several challenges are facing: 1) Events in the specific financial domain are comparatively concentrated in nature. This concentration of event types can lead to an imprecise demarcation between events; 2) The process of linking events can be challenging due to several companies sharing the same short name, as well as the prevalence of subsidiary corporations among most companies; 3) Linked events can be very sparse. For example, certain companies or CEOs may seldom appear in event tuples.

To extract event tuples, common open-domain knowledge extraction methods like OpenIE have been used in previous studies. For instance, Deng et al.~\cite{deng2019knowledge} used OpenIE\footnotemark[5] 
\footnotetext[5]{\url{https://sraf.nd.edu/loughranmcdonald-master-dictionary/}} to generate initial event tuples and then eliminated redundant words, such as adjectives and adverbs, from the set of generated event tuples to ensure their quality. In addition to this, Wang et al.~\cite{wang2021global} developed an event detection model specific to the Chinese financial market, which constructs a financial event graph to aid in event classification. Moreover, since event graphs are constructed for specific financial domains, the types of events tend to be relatively centralized, resulting in fuzzy boundaries between events. Wang et al.~\cite{wang2021global} addressed this challenge by incorporating prior knowledge that imposes a layer of restriction on event detection when classifying financial events.

The second step aims to connect the items in the event tuple to the graph. Entity disambiguation (ED)~\cite{ganea2017deep} and entity linking (EL) technologies~\cite{sil2013re} are used to create a sub-graph from the generated event tuples. As the information in event tuples may be sparse and lack diversity, Deng et al.~\cite{deng2019knowledge} enriched the linked event knowledge with one-hop relations extracted from a commonly open-source knowledge graph, like Freebase~\cite{bollacker2008freebase} and Wikidata~\cite{vrandevcic2014wikidata}.

\begin{table*}[!t] \scriptsize
\renewcommand{\arraystretch}{1.15}
\caption{Summaries of Methods Applied for Knowledge Acquiring}\label{acquiring}
\centering
\resizebox{\textwidth}{!}{
\begin{tabular}{c|l|c|c}
\hline
\bfseries Knowledge &   \bfseries Data Sources  & \bfseries Methods &\bfseries Ref.\\
\hline
Descriptive features
& Financial news & BoW~\cite{groth2011intraday}/OpinionFinder~\cite{wilson2005opinionfinder} & Schumaker et al.~\cite{schumaker2012evaluating} \\
& Corporate disclosures & Chi-Squared + TF-iDF & Groth et al.~\cite{groth2011intraday}\\
 & Corporate announcements & N-Gram/N-word combination & Hagenau et al.~\cite{hagenau2013automated}\\ 
 
\hline
Sentiment scores
& Social media & Logistic Regression/SVM & Gupta et al.~\cite{gupta2020sentiment}\\
& Social media & Sentiment dictionary & Wang et al.~\cite{wang2018combining}\\
& Financial news & LDA~\cite{blei2003latent} & Chen et al.~\cite{chen2018incorporating}\\
& Financial news & Sentiment dictionary\footnotemark[5] & Li et al.~\cite{li2018stock}\\

& Financial news & Correlation matrix & Liu et al.~\cite{liu2019combining}\\
& Social media + Financial news & Stanford NLP~\cite{socher2013recursive} & Khan et al.~\cite{khan2020stock}\\


\hline
Semantic features
& Financial news & Word2vec~\cite{kilimci2020efficient} & Lee et al.~\cite{lee2017predict}\\
& Financial news & Word2Vec + Attention score & Hu et al.~\cite{hu2018listening}\\
& Social media + Financial news & Word2Vec/GloVe~\cite{kilimci2020efficient}/FastText  & Kilimci et al.~\cite{kilimci2020efficient}\\

& Financial news & ELMo~\cite{huang2022news} & Huang et al.~\cite{huang2022news}\\
& Social media & BERT~\cite{devlin2018bert} & Chen et al.~\cite{chen2022gated}\\
& Financial news & BERT & Chen et al.~\cite{chen2021stock}\\
& Financial news & NLTK~\cite{Steven2006nltk} + Fin-BERT~\cite{araci2019finbert}  & Sonkiya et al.~\cite{sonkiya2021stock}\\
& Financial news & Chinese-RoBERTa~\cite{zhao2019uer} & Ma et al.~\cite{ma2023multi}\\



\hline
Audio features
& Earning Conference Calls dataset & Praat~\cite{boersma2001speak} & Qin et al. ~\cite{qin2019you}\\
& Earning Conference Calls dataset & Praat & Yang et al.~\cite{yang2020html}\\

\hline
Homogeneous 
& Historical prices & Correlation matrix  & Li et al.~\cite{li2021modeling}\\
inter-stock 
& Historical prices & Correlation matrix & Sarmah et al.~\cite{sarmah2022learning}\\
correlations 
& Historical prices & Correlation matrix & Chi et al.~\cite{chi2010network}\\
& Historical prices & Correlation matrix + PMFG~\cite{tumminello2005tool} & Yan et al.~\cite{yan2020development}\\
& Historical prices & Correlation matrix + MST~\cite{mantegna1999hierarchical} & Pozzi et al.~\cite{pozzi2013spread}\\
& Historical prices & Correlation matrix & Li et al.~\cite{li2019portfolio}\\
& Historical prices & Correlation matrix + MST & Aste et al.~\cite{aste2010correlation}\\
& Wiki company-based relations & Explicit information & Sawhney et al.~\cite{sawhney2020deep}\\
& Industry information & Explicit information & Ying et al.~\cite{ying2020time}\\

\hline
Heterogeneous
& Financial news & BERT + BiLSTM + Multi-head attention & Cheng et al.~\cite{cheng2020knowledge}\\
inter-stock 
& Historical prices  & Correlation matrix & Xiang et al.~\cite{xiang2022temporal} \\
correlations
& Financial news + Financial reports & Human-interacted OpenNRE~\cite{han-etal-2019-opennre}  & Zhu et al.~\cite{zhu2023hit}\\

\hline

High-order 
& Historical prices + Industry information & Explicit information + Correlation matrix & Huynh et al.~\cite{huynh2023efficient}\\
inter-stock
& Industry information & Explicit information & Li et al.~\cite{Li2022hyper}\\
correlations
& Industry information + Wiki company-based relations\footnotemark[2] & Explicit information & Sawhney et al.~\cite{sawhney2021stock}\\

\hline
Exter-stock
& Fund manager’s portfolio information & Explicit information & Li et al.~\cite{li2019individualized}\\
interactions 

& Mutual funds’ stock holdings information & Explicit information & Lu et al.~\cite{lu2018herding}\\
& Common shareholders information & Explicit information & Bisarya et al.~\cite{bisarya2022stock}\\
& Industry + Wiki company-based relations & Explicit information & Feng et al.~\cite{feng2019temporal}\\
& Industry + Main business of stocks & Explicit information + Correlation matrix & Xu et al.~\cite{xu2021hist}\\
& Industry information & Explicit information & Cardoso et al.~\cite{cardoso2022learning}\\
& Financial analysts + Directors information & Explicit information & Creamer et al.~\cite{creamer2009link}\\

\hline

Linked events
& Financial news + Open source KGs  & Open IE + EL~\cite{sil2013re} & Deng et al.~\cite{deng2019knowledge}\\
& Financial news & BERT & Wang et al.~\cite{wang2021global} \\ 

& Executives information + Company relations\footnotemark[7] 
 & Explicit information   & Zhao et al.~\cite{zhao2022stock} \\
& + Financial news +  Historical prices  & + Attention score & \\
\hline
Temporal patterns
& Financial news + Historical indicators  & Wikipedia2Vec\footnotemark[8]
+ Dot-product attention & Ang et al.~\cite{ang2022learning}\\
& + Open source KGs\footnotemark[9]& & \\

& Historical prices & Daily correlation matrix & Xiang et al.~\cite{xiang2022temporal}\\
& Quarterly bond-fund holding data  & Explicit information & Zhou et al.~\cite{zhou2022temporal}\\
& Financial news + Financial reports & OpenNRE + DART~\cite{lin2018domain} & Zhu et al.~\cite{zhu2023hit}\\
\hline

\end{tabular}}
\vspace{-3mm}
\end{table*}

\footnotetext[7]{\url{https://github.com/dair-iitd/OpenIE-standalone}}
\footnotetext[8]{Four types of company relations from a publicly available API, Tushare: \url{https://tushare.pro/}}
\footnotetext[9]{\url{https://wikipedia2vec.github.io/wikipedia2vec/}}
\footnotetext[10]{\url{https://blog.gdeltproject.org/gdelt-global-knowledge-graph/}}

\subsubsection{Temporal patterns}
In the real market, the relationship between companies will change over time, resulting in the importance of the temporal patterns knowledge to represent the stock market volatility. The temporal knowledge of the stock market can be modeled in dynamic graphs. The edges in the dynamic graph can be inserted or deleted at any timestep. 

\begin{definition}{Temporal patterns network.} The discrete-time dynamic graph is defined as a discrete sequence of static graphs, which can be denoted as $\mathbb{G} = \{\mathcal{G}^1, \mathcal{G}^2, \mathcal{G}^3, ..., \mathcal{G}^T\}$, and each $\mathcal{G}^t$ is a static snapshot graph at timeslice $t$.
\end{definition}
In the above definition, $\mathcal{G}^1$ is the initial state of a dynamic graph, and each graph $\mathcal{G}^t(t > 1)$  evolves from the previous graph $\mathcal{G}^{t-1}$. 
There are three challenges when acquiring spatiotemporal knowledge: irregularity of external knowledge in time, sparsity issues in snapshot graphs due to the low event frequency of some companies, and duplication and conflict problems when updating dynamic graphs.
Researchers need to sample data at regular intervals, such as daily~\cite{ang2022learning,xiang2022temporal}, monthly~\cite{zhou2022temporal}, or quarterly~\cite{zhou2022temporal}. This allows them to create dynamic graphs that reflect the changes in stock prices over time. The use of historical stock prices can simultaneously capture the temporal information of the stock market. It can also address the issue of irregular time-series data, as stock prices inherently represent a regularly recorded time-series dataset. As a result, Xiang et al.~\cite{xiang2022temporal} developed a daily multi-relation graph to represent positive and negative correlations between stocks, based on a series of timestamps $T={ t_1,t_2,...,t_n}$ that indicate each trading day.

Zhou et al.~\cite{zhou2022temporal} utilized a dynamic bipartite graph to present evolved investment interactions between bonds and institutional funds. Similarly, Ang et al.~\cite{ang2022learning} leveraged a public open-domain Global Knowledge Graphs (GCK)\footnotemark[8] dataset to construct a daily dynamic graph focusing on stock companies in NY and NA. To address the sparsity issue in each snapshot graph, they extracted implicit inter-company relations to enrich the GCK. They applied a dot-product attention layer to calculate the stock correlation. The input values are stock representations learned from financial news and historical indicators.

For the third challenge, Zhu et al.~\cite{zhu2023hit} proposed a dynamic financial KB construction pipeline, which uses a human-interacted distant supervised method to extract evolved relations. Importantly, they highlighted the challenges of duplication and conflict resolution for acquiring high-confidence dynamic knowledge and proposed corresponding dynamic graph updating approaches.

\section{Knowledge Incorporation} \label{sec:inc}

  
When incorporating financial knowledge into stock price prediction tasks, it is crucial to focus on three key factors: 1) understanding the dependencies between external knowledge and stock prices; 2) considering the duration of knowledge impact; 3) integrating heterogeneous knowledge. This section introduces existing works that effectively leverage the acquired knowledge (as described in Sec.~\ref{sec:acq}), and provides an overview of different fusion methods for combining stock historical price data with external knowledge.


\subsection{Non-graph-based Knowledge}

\subsubsection{Descriptive features}
Previous studies~\cite{alostad2015directional, luss2015predicting} have primarily treated descriptive features as static features with high-dimensional and sparse characteristics. In these studies, SVM was applied with inputs such as BoW vectors~\cite{luss2015predicting} or document N-Gram matrices~\cite{alostad2015directional}. However, the absence of context information is limited to the usage of this knowledge in recent research.

\subsubsection{Sentiment features}

Sentiment features can be used as a historical technical indicator since most works processed this to a time-series data in the knowledge acquisition process. Prediction models include machine learning models (e.g., SVM~\cite{li2018stock}), RNN-based models (e.g., LSTM~\cite{xu2019stock}, GRU~\cite{liu2019combining}), Transformer-based models (e.g., BERT~\cite{sonkiya2021stock}), and ensemble learning~\cite{wang2018combining}. The considerable influx of user comments pertaining to stock markets on a daily basis contributes to sentiment scores that encompass a high-dimensional feature space. Several works aggregated the sentiment scores of all users on a daily basis to derive the daily overall sentiment~\cite{gupta2020sentiment,khan2020stock}. Incorporating complete sentiment features into prediction models is a significant challenge. To address this challenge, Wang et al.~\cite{wang2018combining} devised an effective solution by employing an ensemble method that combines multiple neural networks.  They adopted a random sampling approach to create several user groups and constructed a sentiment feature subspace for each group. Subsequently, DNNs were trained within each subspace and then combined through the ensemble algorithm. This ensemble approach enhanced the overall prediction performance by leveraging the collective knowledge obtained from diverse user sentiments.

In addition, to tackle the challenge of missing values in sentiment analysis, one potential solution is to integrate graph-based knowledge into the sentiment score computation. By leveraging the interconnected relationships represented in the knowledge graph, the sentiment score of a particular stock can be extended to its associated or related stocks. For instance, Liu et al.~\cite{liu2019combining} multiplied daily sentiment matrixes with a static stock correlation matrix. The final sequence of the sentiment matrix $\mathcal{B} = \{B_1, B_2, ..., B_t\}$ can then be concatenated with other time series features as the input of the RNN-based model.


\subsubsection{Semantic features} 

In comparison to the aforementioned knowledge sources, semantic features offer the advantage of lower dimensionality while providing more intricate contextual information. To establish a temporal alignment between news articles and stock prices, prior researchs~\cite{lee2017predict} have commonly adopted a data merging strategy. Specifically, semantic vectors and stock price features from the same day are concatenated. This simple approach allows for the seamless integration of semantic features with numerical stock data.

However, news articles can have varying impacts depending on their release dates, and different news within the same day may exert different influences. To address these challenges, Huang et al.~\cite{huang2022news} generated two separate embedding matrices to represent news from yesterday and news within the last 30 days. Then, they employed scaled dot-product attention to capture how these news embeddings influence the stock movement of the current day. Moreover, Hu et al.~\cite{hu2018listening} tackled both challenges. Initially, they introduced a news-level attention mechanism for aggregating news articles published on the same day. In their approach, the representation of news on the $t$-th day is obtained by summing the embeddings of individual news articles, each weighted by its respective normalized attention weight. Subsequently, Hu et al.~\cite{hu2018listening} employed a bidirectional Gated Recurrent Unit (bi-GRU) model to capture the dependencies within the news representation sequence denoted as $V$. Unlike using the final state of the GRU as the output, they leveraged the latent state of each date to generate a new state sequence denoted as $V'=[v'{t-T+1}, v'{t-T+2}, ...,v'_{t}]$, where $T$ represents the number of past days considered. To further enhance the temporal information, an attention layer was implemented to weigh the importance of different dates.

\subsubsection{Audio features} 

To align audio recordings and corresponding transcripts, researchers~\cite{yang2020html, qin2019you} generate sentence-level audio and text embeddings. These embeddings, treated as sequence data, are then fed into deep learning models to capture dependencies with historical stock prices. Qin et al.~\cite{qin2019you} used the RNN-based method, a bidirectional LSTM (BiLSTM), to learn the temporal dependencies. Yang et al.~\cite{yang2020html} used the transformer-based method, which demonstrated better performance compared to RNNs. Since earnings calls are held every quarter, which presents challenges in aligning with the stock price series. The previous approach to incorporating stock price prediction involves forecasting the volatility of the stock price within n days after the earnings conference call.

\subsection{Graph-based Knowledge}
Incorporating graph knowledge into existing prediction models can be a challenging task due to various factors. Firstly, financial graphs can be large-scale, making it difficult to process and analyze the data effectively. Secondly, the entities and relationships in financial graphs can be heterogeneous, leading to complexity in modeling and interpretation. Additionally, the implicit influence of event information on financial entities, such as stock-dependent influence and cross-stock influence~\cite{xu2021rest}, can further complicate the integration of graph knowledge into prediction models. Finally, dynamic knowledge also poses a challenge as financial graphs are constantly evolving, and traditional static modeling techniques may not be sufficient to capture the changes accurately.


Inspired by graph embedding methods, there has been a surge in the number of studies that utilize them to incorporate financial graph-based knowledge into stock price prediction models. These methods aim to obtain low-dimensional vector representations of nodes in the graph while preserving both the graph topology and node information. Two main types of graph embedding learning methods have been used in these studies: \textbf{random walk-based methods} and \textbf{graph neural network (GNN)-based methods}. In addition, graph-based knowledge-enhanced stock prediction methods typically incorporate not only graph-based knowledge but also other sources of information, such as text features and historical prices, which will be introduced in the next subsection. Two main graph embedding techniques will be mentioned in the first part (Sec.\ref{Homo-incorporating}). Since there are challenges that arise when these general methods are used in incorporating different kinds of financial graph-based knowledge, the corresponding solutions will be introduced separately for each type of knowledge.

\subsubsection{Homogeneous inter-stock correlations}\label{Homo-incorporating}

By leveraging interdependencies and correlations between stocks to improve stock price prediction accuracy. The use of graph embedding learning methods is particularly advantageous as it allows for the efficient handling of large-scale financial graphs and facilitates the extraction of meaningful features from complex financial data.




\textbf{Random walk-based method.} 
In accordance with the concept of representational learning and word embedding, existing methods~\cite{grover2016node2vec,perozzi2014deepwalk,tang2015line,yang2015network} are proposed to learn graph embedding based on random walks, which treat walks as the equivalent of sentences. Random walk-based graph embedding methods first generate a sequence of nodes from the graph and then apply the Skip-Gram model~\cite{mikolov2013efficient} to the generated walks. They can preserve the graph structure information and further take into account the context path information of the target node. The methods believe that two nodes linked by a strong link should be similar and close in place. Chen et al.~\cite{chen2018incorporating} apply several random walk-based graph embedding methods (DeepWalk~\cite{perozzi2014deepwalk}, LINE~\cite{tang2015line}, and node2vec~\cite{grover2016node2vec}) to identify the most related corporation for the target stock by calculating the similarity of node embedding. Then they took the average of the top K-related stock representation, denoted as $X_i^{\prime}$, and concatenated it with the target node embedding, denoted as $X$. This merged embedding vector, denoted as $X X_i^{\prime} \in \mathbb{R}^{2d\times t}$, is fed into an LSTM-based encoder, where $d$ is the dimension of node embedding and $t$ is the number of trading days.



Although random walk-based methods have been demonstrated its success in learning latent representations of target stock based on similar corporations, the lack of parameter sharing between nodes and the inability to handle dynamic or unseen graphs limit their usefulness in more complex graph-related tasks~\cite{zhou2020graph}. Recent works have addressed these limitations by using GNN-based methods to learn node representations, which enable more efficient and adaptable graph modeling. By utilizing GNN-based methods, it is possible to capture the correlations and dependencies between the nodes more effectively, making it a more practical approach to graph-related tasks. Therefore, recent works have been performed to learn the graph representation by GNN-based methods. 

\textbf{GNN-based method.}
GNN is the abbreviation for graph neural network, which refers to the model applied by deep learning neural networks on graphs. In terms of the way of propagation, GNN~\cite{zhou2020graph} can be classified into graph convolutional network (GCN)~\cite{kipf2016semi}, graph attention network (GAT)~\cite{velivckovic2017graph} and etc. GCN and GAT are the most prevalent models employed by existing methods. GCN applies graph Laplacian to generate the feature maps inspired by the CNN operation. Besides the GCN, GAT applies the attention mechanism to determine the importance of neighbor nodes during the aggregation. Inputs of GNNs contain \textit{a graph adjacency matrix} and \textit{a feature matrix}. The adjacency matrix, $A^r \in \mathbb{R}^{|\mathcal{V}|\times|\mathcal{V}|}$, can represent the graph structure and its connection information. The adjacency matrix is labeled by the graph vertices, with a 0 or 1 in position $(v_i,v_j)$ according to whether the edge $e_k = (v_i,v_j)$ exists. The feature matrix, $\textbf{X}\in \mathbb{R}^{|\mathcal{V}|\times d}$, denotes a set of node features, where $d$ is the dimension of the embedding. The initialization of the feature matrix is commonly done randomly. However, it can also be initialized using features learned from other input data sources, such as historical prices and textual data, to incorporate multiple types of inputs.


Chen et al.~\cite{chen2018incorporating} also proposed a hybrid stock price movement prediction model. They applied GCN to incorporate information from the whole graph network and an LSTM-based encoder to process the historical stock price feature. The input feature matrix of the GCN model is obtained from the LSTM layer. The GCN layer can be represented as follows:

\begin{equation}
H^l=\operatorname{\it \sigma}\left(\tilde{D}^{-\frac{1}{2}} \tilde{A} \tilde{D}^{-\frac{1}{2}} H^{l-1} W^{(l-1)}\right) 
\end{equation}
where $\widehat{A} = A + I_N$ is the adjacency matrix of the graph and $\tilde{D}_{i i}=\sum_j \tilde{A}_{i j}$~\cite{kipf2016semi}.
They utilized three GCN layers to reconstruct the historical sequence features by incorporating the corporation connection graph.

Sawhney et al.~\cite{sawhney2020deep} devised a multipronged attention network for stock forecasting (MAN-SF) that integrates a GAT and simultaneously learns from various sources: historical sequence prices, social media textual knowledge, and the homogeneous stock graph. To incorporate non-graph-related knowledge, they introduced a price encoder (PE) and a social media information encoder (SMI). The PE utilizes a GRU to capture sequential dependencies across trading days, followed by temporal attention to aggregate the hidden representation from the GRU. On the other hand, the SMI is used to process social media knowledge, where it generates sentence-level tweet embeddings using universal sentence encoders (USE)~\cite{cer2018universal}. The above encoder produces price feature $q_t$ and social media feature $c_t$. For the purpose of learning pair-wise interactions, multi-modal knowledge, namely historical and textual knowledge, is merged using a bi-linear transformation.
\begin{equation}
x_t=\mathcal{B}\left(c_t, q_t,\right)=R e L U\left(q_t^T W c_t+b\right)
\end{equation}
The input feature matrix for the GAT layer in MAN-SF is a set of node feature $\mathbf{h} = \{x_1,x_2,...,x_{|\mathcal{V}|}\}$. The GAT propagates the fused feature based on the connections on the stock homogeneous graph.
The normalized attention coefficients $\alpha_{i,j}$ reflect the significance of the connections among stocks $i$ and $j$, which can be calculated by:
\begin{equation}
\alpha_{i j}=\frac{\exp \left(\operatorname{LeakyReLU}\left(a_w^T\left[W x_i \oplus W x_j\right]\right)\right)}{\sum_{k \in N_i} \exp \left(\operatorname{LeakyReLU}\left(a_w^T\left[W x_i \oplus W x_k\right]\right)\right)}
\end{equation}
MAN-SF uses GAT to update the feature matrix by weighting and aggregating the features of neighbor nodes based on the attention coefficients $\alpha_{i,j}$. 


\subsubsection{Heterogeneous inter-stock correlations}

Knowledge incorporating methods for the heterogeneous inter-stock correlation is required to learn separate embeddings for different types of relations. To formulate the heterogeneous relations, most existing works~\cite{feng2019temporal,xu2021rest} set the input of the incorporating model as \textit{a set of adjacency matrices} $\mathcal{A}\in \mathbb{R}^{|\mathcal{V}|\times|\mathcal{V}|\times k}$, where $|\mathcal{V}|$ is the number of nodes, and $k$ is the dimension of the feature vector. For each relation $r \in \mathcal{R}$, there is an adjacent matrix $A^r \in \mathcal{A}$, where $A^r \in \mathbb{R}^{|\mathcal{V}|\times|\mathcal{V}|}$. For instance, there are three stocks ($v_i$, $v_j$, and $v_k$) and two relationships named "supplier" and "same\_industry". Suppose stock $v_i$ is a supplier of stock $v_j$, while stock $v_i$ belongs to the same industry as stock $v_k$. Their relationships can be encoded as two different vectors: $\mathcal{A}_{ij} = [1,0]$ and $\mathcal{A}_{ik} = [0,1]$~\cite{feng2019temporal}. After that, by using a learnable function that takes in different relationship vectors for distinct stock pairs, we can enable the embedding propagation process to account for the topological structure of the relationship graph and the meaning of the relationships. This approach allows us to capture the complex interdependencies between the stocks and their relationships.

Xu et al.~\cite{xu2021rest} constructed a multi-relational graph to capture the impact of events on other related stocks (known as a cross-stock influence). The original GCN mode cannot distinguish the impacts of different relations. They developed a relational GCN propagation layer that enables the diffusion of event information from associated stocks within the multi-relational stock graph.
\begin{equation*}
H_t^l=\sum_{r \in \mathcal{R}} \tilde{A}^r W^r H^{l-1}_t = 
\sum_{r \in \mathcal{R}} D^{-\frac{1}{2}} A^r D^{-\frac{1}{2}} W^r H^{l-1}_t
\end{equation*}
where $H_t^l$ is the effect of the event information on date $t$. Unlike the GNN-based model we discussed in the homogeneous graph, Xu et al.~\cite{xu2021rest} presented a distinct approach. They included an adjacent matrix labeled as $A^r$ and a mapping matrix named $W^r$, both of which are dedicated to a specific relationship $r. $ The effect of the event propagated using relationship $r$ is demonstrated by $W^r H^{l}_t$. As a result, the relational GCN propagation layer has the ability to comprehend how to propagate event effects under varying relationships.

Feng et al.~\cite{feng2019temporal} proposed a temporal graph convolution network that contains two propagation for relational embedding and one propagation for time-aware embedding except for the GCN layer. To perform relational embedding, they initially utilized a uniform embedding propagation method to capture the effects of all other stocks that have relations with stock $v_i$. This approach considers all types of relationships, allowing the embedding $e_i$ to encode $e_j$ if the condition $sum(\mathcal{A}_{ij}) > 0$ is satisfied. Subsequently, Feng et al. utilized a weighted embedding propagation method to capture the varying effects that arise from different relationships between two stocks. The weighted embedding propagation can be formulated as follows:

\begin{equation}
\overline{\mathbf{e}}_{i}=\sum_{\left\{j \mid \operatorname{sum}\left(\mathcal{A}_{ij}\right)>0\right\}} \frac{g\left(\mathcal{A}_{ij}\right)}{d_j} \mathbf{e}_{j},
\end{equation}
where $d_j$ is the number of stocks metting the condition $sum(\mathcal{A}_{ij}) > 0$ and $g\left(\mathcal{A}_{ij}\right)$ is a relation-strength mapping function. The purpose of this function is to capture the interaction between two stocks, either explicitly or implicitly. The explicit modeling considers the similarity between the two stocks and the importance of the relations. 

\subsubsection{High-order inter-stock correlations}

Current works rely heavily on the \textit{hypergraph convolution mechanism}~\cite{yadati2019hypergcn}. These methods typically first capture the long-term temporal features from stock time-series data utilizing LSTM and temporal attention and feed the features into the hypergraph convolution network. To effectively bridge the gap between the temporal attention and hypergraph convolutions, Sawhney et al.~\cite{sawhney2021stock} applied a hypergraph attention mechanism that learns to dynamically assign weight to each hyperedge based on its corresponding stock temporal feature.

However, the hypergraph becomes more complex when dealing with the stock market. The high-order inter-stock correlation knowledge~\cite{Li2022hyper}, which is solely obtainable from industry categories, can be represented by a hypergraph with 104 hyperedges. As a result, computing the hypergraph Laplacian using the original Fourier basis~\cite{yadati2019hypergcn} during the factorization process can be prohibitively expensive. To address this issue, Huynh et al.~\cite{huynh2023efficient} have proposed the use of the Wavelet basis~\cite{xu2019graph}, which is much sparser than the Fourier basis and allows for more efficient computation.

\subsubsection{Exter-stock interactions}

The challenge of incorporating bipartite graphs lies in their heterogeneous nature, as the entities involved are often located in different feature spaces. Aggregating information across these different spaces presents difficulties due to the heterogeneity of attributes. Nonetheless, in this financial-specific task, the primary objective of incorporating bipartite graphs is to facilitate the aggregation of concept information with respect to the relevant stocks. 

To address this challenge, some existing works~\cite{bisarya2022stock,creamer2009link,lu2018herding} projected bipartite graphs into homogeneous graphs while retaining information on interactions in the original graph. Only the stock entity is retained in the projection-based method. Bisarya et al.~\cite{bisarya2022stock} set the weight of link $e_{i,j}$ using the average value of weights of links $e_{i,k}$ and $e_{j,k}$, where $v_i, v_j \in \mathcal{V}_A$ and $v_k \in \mathcal{V}_B$. Creamer et al.~\cite{creamer2009link} set weights for the projected homogeneous graph as the count number of common neighbors. Lu et al.~\cite{lu2018herding} constructed a directed investor-stock company bipartite graph which calculated the projected weights by $w_{ij} = \sum_{k\in C_{ij}}w_{ki}$ and $w_{ji} = \sum_{k\in C_{ij}}w_{kj}$, where $C_{ij}$ is the common neighbors of stock nodes $v_i,v_j$ in the raw bipartite graph. Following the projection step, the process of incorporation can be performed similarly to the methods used for the inter-stock correlation knowledge.

Other methods enable the retention of information from both sets of nodes as well as the weighted edges between them~\cite{li2019individualized, xu2021hist,zhou2022temporal}. To incorporate the heterogeneous knowledge, some methods leverage financial domain knowledge to determine the \textit{meta-path} and aggregate interaction information of connected concept nodes for the target stock node. Li et al.~\cite{li2019individualized} employed a fund-stock bipartite graph and observed that stocks linked by the same funds tend to have similar characteristics. To exploit this observation, they proposed a two-step random walk method that generates the random walk sequence exclusively with stock nodes connected by the same fund node. Furthermore, Xu et al.~\cite{xu2021hist} utilized an attention mechanism to aggregate the representations of the same concepts to the stocks. They stated that since different concepts have varying importance for the stocks, the attention mechanism is utilized to learn the relevance of each concept to a stock. Additionally, considering the dynamics of the stock market, the stock price for a particular stock can be impacted by different concepts at different times. Cardoso et al.~\cite{cardoso2022learning} proposed a novel dynamic soft clustering algorithm to capture implicit knowledge according to historical stock prices.


In support of the bipartite graph, Lu et al.~\cite{lu2018herding} analyzed the real market crash data and the knowledge presented in the investor-stock bipartite graph they constructed. They concluded that too-connected stocks are more stable in a crash, and the drop in their price would have a significant effect on the related stocks. This supports incorporating interaction knowledge does provide brilliant insight into the stock market.

\subsubsection{Linked events}
Linked event knowledge incorporation requires preserving the semantic information of the event and aggregating them to the representations of relevant stocks that are impacted by the event. To address the challenge,  Deng et al.~\cite{deng2019knowledge} employed TransE to learn the structural information of the event graph. Additionally, they proposed a multi-channel method to represent the event by graph linking $\mathbf{X}_l$, graph context $\mathbf{X}_c$, and word vectors $\mathbf{X}_w$. The concatenation of $\mathbf{X}_l$, $\mathbf{X}_c$ and $\mathbf{X}_w$ as the financial event representation.
Then they concatenated the stock price vector with the event vector and fed the resulting embedding into a temporal convolution network-based prediction model. Furthermore, they interpret the event effects by computing event contribution towards or against the predicted class in a binary-class classifier. For $ i_{th}$ event tuple, the effect can be defined by:
\begin{equation}
    Effect_{i}=\theta_i C_i
\end{equation}
where $\theta_i$ is the weight coefficient and $C_i$ is instance value of events.


The duration of event impact on stock prices also poses a challenge when incorporating event knowledge into stock price prediction models. One simplest solution, as proposed by Ding et al.~\cite{ding2016knowledge}, involves utilizing events from one day to predict the stock price of the next day. However, the influence of an event on stock prices may extend beyond a single day~\cite{xu2021rest}. Ding et al.~\cite{ding2015deep}, categorized events into long-term, medium-term, and short-term events. For learning long-term and medium-term events, they employed a one-dimensional convolution function and max pooling to capture concentrated information. In the case of short-term events, event embeddings were directly averaged to preserve the relevant information. However, when utilizing a one-dimensional convolution function to learn a whole sequence of event features, the issue of information leakage from future data arises. This leakage can significantly impact the accuracy of test set results. To overcome this challenge, Deng et al.~\cite{deng2019knowledge} proposed a Temporal Convolutional Network (TCN) with causal convolution. In this approach, the convolution is performed at time t, ensuring that only elements up to time t are convolved.

In addition, there have been several attempts to incorporate event information that combines pre-generated event embeddings with either homogeneous or multi-relational stock graph embeddings. One such study was conducted by Cheng et al.~\cite{cheng2020knowledge}, who integrated event embeddings, graph embeddings, and relation embeddings using a multi-source attention layer to capture their distinct contributions. Their approach integrated the event embedding for the related node in a multi-relational graph. However, a limitation of these methods is that they typically did not generate explicit event nodes in the graph, which limits the ability of the model to capture the underlying dynamics of the market fully. Nonetheless, the approach proposed by Cheng et al.~\cite{cheng2020knowledge}  outperformed traditional models that did not utilize event information, underscoring the potential benefits of incorporating both event and inter-stock correlation in the financial analysis.


\begin{table*}[!t]
\renewcommand{\arraystretch}{1.3}
\caption{Summaries of Models Applied to Incorporate the External Knowledge into Stock Price Prediction}\label{incor}
\centering
\resizebox{\textwidth}{!}{

\begin{tabular}{ccccl}
\toprule
\bfseries Knowledge &   \bfseries Incorporation model  & \bfseries With time-series? & \bfseries Fusion method &\bfseries Ref.\\
\hline
Descriptive features 
& RNN-boost & Y & Parallel (timestep-wise) & Chen et al.~\cite{chen2018leveraging}\\
& Bootstrap-DL & Y & Parallel (timestep-wise)  & Wang et al.~\cite{wang2018combining}\\
\hline 
Sentiment scores

& GAN & Y & Parallel (timestep-wise) & Sonkiya et al.~\cite{sonkiya2021stock}\\
& GCN + BiLSTM & Y & Parallel (timestep-wise)  & Ma et al.~\cite{ma2023multi}\\
& GRU & Y & Parallel (timestep-wise) & Liu et al.~\cite{liu2019combining}\\

\hline
Semantic features & CNN/RNN-based & N & - & Kilimci et al.~\cite{kilimci2020efficient}\\
& CNN + LSTM & Y & Parallel (timestep-wise)  & Lee et al.~\cite{lee2017predict}\\
& Attention + BiGRU & N & - & Hu et al.~\cite{hu2018listening}\\
& LSTM & N & - & Haung et al.~\cite{huang2022news}\\
& Transformer-based & Y & Parallel (last-stage) & Chen et al.~\cite{chen2022gated}\\
& RNN-based & N & - & Chen et al.~\cite{chen2021stock}\\


\hline
Audio features & Transformer-based & N & - & Yang et al.~\cite{yang2020html}\\
& BiLSTM & N & - & Qin et al.~\cite{qin2019you}\\

\hline
Homogeneous 
 
& Random walk-based (Node2vec~\cite{grover2016node2vec}) &  N & - & Sarmah et al.\cite{sarmah2022learning} \\
inter-stock correlations 
& GCN-based & N & - & Li et al.~\cite{li2021modeling} \\
& GAT-based &  Y & Sequential  & Sawhney et al.~\cite{sawhney2020deep} \\
& Random walk-based methods/ GCN-based  & Y & Sequential & Chen et al.~\cite{chen2018incorporating} \\
 \hline
Heterogeneous 
& Random walk-based (Node2vec~\cite{grover2016node2vec}) & Y & Parallel (last-stage) & Saha et al.~\cite{saha2021stock}\\
inter-stock correlations 
& GCN-based & Y & Sequential & Feng et al.~\cite{feng2019temporal}\\
& GCN-based & Y & Sequential & Matsunaga et al.~\cite{matsunaga2019exploring}\\
 & GCN-based & Y & Sequential & Xu et al.~\cite{xu2021rest}\\ 
 & Random walk-based (TriDNR~\cite{pan2016tri}) & N & - & Cheng et al.~\cite{cheng2020knowledge} \\
\hline
High-order & HyperGCN & Y & Sequential & Sawhney et al.~\cite{sawhney2021stock}\\ 
inter-stock correlations & HyperGCN using Wavelets & Y &Sequential & Huynh et al.~\cite{huynh2023efficient}\\ 
\hline
Exter-stock correlations
 & Graph statistics + Logitboost  & N & - & Creamer et al.~\cite{creamer2009link}\\

& Node2vec + LSTM & Y &Parallel (last-stage) & Bisarya et al.~\cite{bisarya2022stock}\\
& GCN-based + LSTM & Y & Sequential & Zhou et al.~\cite{zhou2022temporal}\\
& GRU + GAT-based & Y & Sequential& Xu et al.~\cite{xu2021hist}\\
& Random walk-based methods & Y & Sequential & Li et al.~\cite{li2019individualized}\\
 
\hline
Linked events 
&  TransE + ConvNets & Y & Parallel (timestep-wise) & Deng et al.~\cite{deng2019knowledge}\\
& TransD + Logistic regression &N & - & Xie~\cite{9670610}\\
\hline

Temporal 
& GCN-based + LSTM & Y & Sequential & Zhou et al.~\cite{zhou2022temporal}\\ 
patterns
& GCN-based + Dynamic SSL & Y & Sequential & Ang et al.~\cite{ang2022learning}\\ 
& GCN-based & Y & Parallel (timestep-wise) & Ma et al.~\cite{ma2023multi}\\

\bottomrule
\end{tabular}}
\vspace{-3mm}
\end{table*}

\subsubsection{Temporal patterns}
Temporal patterns are modeled in dynamic graphs. In the stock price prediction domain, existing works typically treat the dynamic graph as a series of static snapshot graphs, each of which represents the graph at a specific time point. According to this issue, researchers have drawn inspiration from the discrete-time dynamic graph neural network (DTDG) to solve the financial analysis problem, which proposed a framework that combines a \textit{static graph encoder} and a \textit{sequential encoder}~\cite{skarding2021foundations}. The static graph encoder is capable of capturing the structure and attribute information of each snapshot graph, while the sequential decoder is responsible for capturing the temporal evolution representation of a series of static snapshot graphs. For example, Zhou et al.~\cite{zhou2022temporal} incorporated knowledge from a bond-fund dynamic bipartite graph using the DTDG framework. They applied GCN to obtain node embeddings for the target stock node in each static graph and then fed them into an LSTM model to learn the temporal features of the sequence. 

However, in the context of financial analysis, the problem often involves multiple input data types and requires the model to capture multivariate features. To tackle this challenge, Ang et al.~\cite{ang2022learning} developed a novel approach that integrates dynamic inter-stock knowledge discovered from public KB, historical prices, and financial news. They constructed a dynamic graph for each kind of input data and utilized GCN as a static graph encoder to aggregate neighbor information for each node in the acquired graphs across their time steps. In order to enhance the model's capacity to learn periodic and non-periodic components within the multivariate features and capture seasonality-based temporal patterns, they incorporated a time vectorization module inspired by time2vec~\cite{kazemi2019time2vec}. This module takes a sequence of time stamps $T_t \in \mathbb{R}^{|V| \times K \times d^{time}}$ as input, which can include the day of the week or the week of the year, with $d^{time}$ representing the dimensions of the captured time-stamp feature. Finally, to ensure consistency across the representations of temporal patterns obtained from multiple dynamic graphs, Ang et al.~\cite{ang2022learning} further proposed a dynamic self-supervised learning (SSL) approach that can effectively handle non-stationary time-series distributions and correlations. 

In addition, the temporal patterns are often irregular and sparse, presenting a challenge for knowledge incorporation models. For instance, Zhou et al.~\cite{zhou2022temporal} have noted that a large number of bonds are traded only a few times during their lifetime in the secondary market, and the news related to a target stock may not appear on a daily basis. This irregularity results in an inadequate number of training samples, which further exacerbates the challenge for the knowledge incorporation model. To overcome this challenge, they introduced a self-supervision model that utilizes the underlying graph structure, rather than target variables, to supervise the learning of representations for all nodes. The model aims to incorporate regularization of the learned node embeddings in order to reconstruct the entire graph topology and edge weights.

\subsection{Time-series and External Knowledge Fusion}

\begin{figure}[t]
    \centering
    \includegraphics[width=\linewidth]{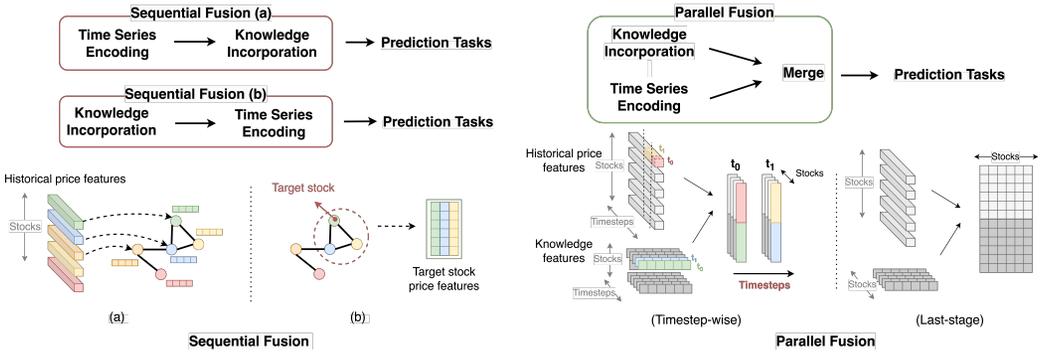}
    \caption{An Illustration of Different Fusion Methods for Time-series Features and External Knowledge Features.}
    \label{fusion}
\vspace{-5mm}
\end{figure}

Stock historical indicators also have been proven to be strong predictors of future trends in stock prices~\cite{xu2021rest}. This time-series data can provides a foundation of temporal information, capturing the sequential patterns and trends in stock prices over time. These indicators involve daily prices and technical factors, including open price, close price, trading volume, moving averages (MA), and relative strength index (RSI), among others. Extensive surveys have highlighted the widespread adoption of combining time-series historical indicators with external knowledge in stock price prediction models. Fusion methods, employed to incorporate both historical price features and external knowledge features, can be classified into two distinct paradigms (shown in Fig.~\ref{fusion}): sequential fusion methods and parallel fusion methods. 


\begin{itemize}
    \item \textit{Sequential fusion:} The order between the incorporation processes of external knowledge and historical prices is sequential. External knowledge incorporation can precede historical price incorporation, and vice versa.
    \item \textit{Parallel fusion:} The order between the incorporation processes of external knowledge and historical prices is parallel. Parallel fusion includes \textit{timestep-wise fusion} and \textit{last-stage fusion}.
\end{itemize}

\subsubsection{Sequential fusion}

The sequential fusion method can be categorized into two groups: 1) The external knowledge incorporation is sequentially followed by the historical prices incorporation; 2)  The historical prices incorporation is sequentially followed by external knowledge incorporation.  

The first approach (shown in Fig.~\ref{fusion} \textit{Sequential Fusion (a)}) is more commonly adopted, especially when incorporating graph-based knowledge. In this approach, the initial input to the knowledge incorporation model is provided by the historical price features. These features are aligned with the corresponding stock node embeddings, facilitating the integration of external knowledge into the model. By utilizing the graph representation, the incorporation model can update the historical price features by considering correlated stock features and structural relations encoded in the graph.
In a work proposed by Sawhney et al.~\cite{sawhney2021stock}, historical indicators are fed into LSTM and attention layers, enabling the capture of temporal features on an individual stock basis. Subsequently, the obtained features are fed into the hypergraph convolution layer, serving as the initialized features for the corresponding stock node. Xiang et al.~\cite{xiang2022temporal} leveraged a multi-head attention transformer to encode the historical prices of each stock across a sequence of trading days. They then flatten the resulting historical price encoding output, denoted as $H\in \mathbb{R}^{|\mathcal{V}|\times T\times d}$, into a two-dimensional matrix represented as $H\in \mathbb{R}^{|\mathcal{V}|\times Td}$. This flattened feature matrix represents a collection of initial embedding vectors for all nodes in the input graph, facilitating message aggregation from neighboring nodes.


The second approach (shown in Fig.~\ref{fusion} \textit{Sequential Fusion (b)}) leverages external knowledge for generating more powerful historical indicators. Li et al.~\cite{li2019individualized} observed that stocks with distinct intrinsic properties exhibit varying levels of affinity or correlation with technical indicators. They proposed a technical trading indicator optimization model (TTIO)~\cite{li2019individualized} to optimize the raw historical price features by stock properties. The stock properties they utilized include stock similarities and stock representations obtained from the knowledge incorporation model. Chen et al.~\cite{chen2018incorporating} incorporated historical price features of relevant stocks into the stock price prediction model, based on this fusion approach. They first applied random walk-based methods like Node2vec~\cite{grover2016node2vec} and LINE~\cite{tang2015line} to learn stock node representations from the graph-based knowledge. By utilizing these learned representations, Chen et al.~\cite{chen2018incorporating} calculated the cosine similarity between stocks, enabling the identification of the most relevant stocks for a given target stock, then combined the historical price features of related stocks with the features of the target stock.

\subsubsection{Parallel fusion}

Parallel fusion methods can be future categorized into two groups: 1) Timestep-wise fusion treats both modalities equally to form a compact input at each time step; 2) Last-stage fusion models each type of knowledge separately and blends them together at the final layer. 

In timestep-wise fusion, the input of the final prediction layer is constructed by merging historical price features and external knowledge features at each timestamp. This approach is commonly used when the external features contain timestamps, such as sentiment scores~\cite{akita2016deep,li2020multimodal} and semantic features~\cite{ma2023multi} obtained from daily news or social media. By incorporating the external knowledge associated with the stock price at each timestep, the model can potentially capture additional insights or patterns that may impact the price behavior. Graph-based temporal patterns knowledge can also be utilized in timestep-wise fusion. Ma et al.~\cite{ma2023multi} employed a multi-source input approach, incorporating historical price indicators, semantic features from news, and homogeneous inter-stock correlation knowledge. They constructed three feature matrices from each knowledge incorporation module, with each feature matrix represented as $X \in \mathbb{R}^{T\times d}$, where $T$ denotes timesteps. These three individual matrices were then concatenated according to the same timestep, creating a combination of daily features. The resulting feature matrix was fed into a classifier to predict stock price movements for the next trading day.

 

Last-stage fusion methods merge the historical price features and external knowledge features as inputs to the final prediction layers. Unlike the timestep-wise fusion methods that consider timesteps, last-stage fusion methods do not explicitly incorporate the temporal aspect. In these methods, the focus is on combining the two sources of information without considering their sequential order, which offers simplicity and efficiency. For instance, in the work of Li et al.~\cite{lil2022flag}, the historical price feature and external knowledge feature obtained from news were concatenated. This concatenated representation was then fed into a linear layer to predict the movement of the stock price. 

Among all the fusion methods, timestep-wise fusion excels at incorporating temporal information. However, external knowledge data may have irregular time series, meaning that their sampling and update frequencies might be inconsistent. Ensuring temporal consistency between external knowledge features and historical price features is crucial.

\section{Datasets of Knowledge-enhanced Stock Price Prediction}\label{sec:data}
Creating a universal benchmark dataset for knowledge-enhanced stock price prediction, capable of accommodating all approaches, is challenging due to the wide range of knowledge types that can be integrated into these models.
Typically, researchers either acquire knowledge themselves or leverage subsets of publicly available open-source knowledge bases (KBs). In our study, we have gathered financial domain-specific KBs that have been utilized in previous works and are accessible to the public. To provide insights into financial domain-specific KBs, we have plotted their statistical information in Table~\ref{KBdata}, followed by a comprehensive analysis. Additional details regarding the data sources for acquiring knowledge mentioned in the paper can be found in Appendix~\ref{app}.

\subsection{Public Financial domain-specific KBs}

\textbf{FR2KG}~\cite{wang2021data} is a Chinese financial knowledge graph that aims to provide the knowledge as well as the raw unstructured texts for the automated KB construction. FR2KG consists of $10$ entity types, $19$ relation types, and $6$ attributes together with $1,200$ unstructured texts of financial research reports as the raw data inputs for extraction. FR2KG is a static KB that is annotated by trained annotators, thus, it has not been updated since August 2021.

\textbf{ShanghaiTech University KG}~\cite{guo2020knowledge} is a financial event KB where events are defined as structured knowledge triple (event argument, event type, event role). Guo et al. initially introduce information from the Chinese National Enterprise Credit Information Publicity System to construct a directed KG with $441,127$ pieces of knowledge with $2$ entity types and $7$ relation types. They further utilize the KG and perform event extraction on the
Chinese financial announcements of the listed companies in China to construct the event KG. Specifically, the event KG still has $2$ event argument types, \textit{person} and \textit{company}, and $6$ event types, which are ``Equity Freeze”, ``Equity Repurchase”, ``Equity Overweight”, ``Equity Underweight”, ``Equity Pledge” and ``Lawsuit”. Moreover, the event KG construction pipeline is four-fold. Firstly, it embeds the entities by their relations into vectors. It secondly uses named entity recognition (NER) to obtain the candidate entities. Thirdly, it obtains the embedding of entities by corresponding to the extracted entities and encodes the embedding. Finally, it uses a series of path-expanding on extracting events.

\begin{table*}[!t]
\renewcommand{\arraystretch}{1.4}
\caption{Statistics of Public Financial Domain-specific KBs (\# denotes "the number of") }\label{KBdata}
\centering
\resizebox{0.85\textwidth}{!}{
\begin{tabular}{cccc}
\toprule
 \bfseries Financial KBs &   \bfseries \# Relations & \bfseries \# Entity Types  &\bfseries \# Relation/Event Types \\
\midrule
 FR2KG~\cite{wang2021data}  & 26,798 & 10 & 19 \\
ShanghaiTech University KG~\cite{guo2020knowledge} & 441,127 & 2 & 7 \\
Chinese RE datasets in Finance~\cite{wu2020creating} & 55,032  & 3 & 6 \\
DuEE-Fin~\cite{han2022duee} & 15,850  & 92  & 13 \\
DCFEE~\cite{yang2018dcfee} & 38,956  &  5 & 4 \\
Doc2EDAG~\cite{zheng2019doc2edag} & 32,040  &  5 & 5 \\
HiDy~\cite{Kquant} & 492,600 &  17 & 34 \\
\bottomrule
\end{tabular}}
\vspace{-5mm}
\end{table*}

\textbf{RE datasets in Finance}~\cite{jabbari2020french,sharma2022finred,wu2020creating} are typical knowledge sources that aim to provide datasets for relation extraction tasks. The RE datasets provide the entities and their annotated relation along with the unstructured text, which can be further integrated into the knowledge triple $(entity_1, relation, entity_2)$. There are RE datasets in different languages. Especially, Wu et al.~\cite{wu2020creating} release a Chinese RE dataset which consists of $6$ relation types ``sue”, ``debt”, ``warrant”, ``asset”, ``asset swap”, ``entrust finance”, and ``pledge” as well as the financial news. Furthermore, there are $3$ entity types: ``company”, ``bank”, and ``securities institution”. The construction can mainly be divided into three steps. The first step is the data resource collection and analysis, which is the unstructured financial news text in this task. The second step is the financial relations definition. Finally, they use POS (Part-of-speech) tagging to do entity recognition annotation and annotate the relations between recognized entities.

\textbf{Event extraction datasets}~\cite{han2022duee,yang2018dcfee,zheng2019doc2edag} aim to provide the dataset for event extraction task, which pre-defines the schema and extracts structured events from unstructured text. Duee-Fin~\cite{han2022duee} is a large-scale benchmark KB labeled by humans which consists of $13$ event types such as ``equity pledge”, ``financing”, ``company listing” and $92$ argument roles. Their dataset construction comprises event schema definition, raw text data collection, and manual annotation. DCFEE~\cite{yang2018dcfee} provides
a labeled document-level event KB along with $2,976$ announcements for the event extraction. The $4$ event types are ``Equity Freeze”, ``Equity Underweight”, ``Equity Overweight”, and ``Equity Pledge”. In addition, the key arguments are ``Name”, ``Pledge Institution”, ``Number of Pledged Stock”, ``Pledging Start Date”, and ``Pledging End Date”. DCFEE constructs the KB via manual annotation on the automatically generated data. Doc2EDAG~\cite{zheng2019doc2edag} is also a labeled document-level event extraction dataset based on ChFinAnn\footnotemark[11]\footnotetext[11]{\url{http://www.cninfo.com.cn/new/index}} documents. There are $5$ event types, where four of them are the same as in DCFEE, and the other is ``Equity Re-purchase”. Moreover, there are $5$ entity mention types mentioned in their paper, which are ``Person”, ``Organization”, ``Date”, ``Share”, and ``Ratio”. The construction of Doc2EDAG is two-fold, which includes document collection and event labeling.

\textbf{HiDy}~\cite{Kquant} is a dynamic financial knowledge base that is organized in a well-formed hierarchy, \textit{Macro}, \textit{Meso}, \textit{Micro}, and \textit{Others}. HiDy is currently one of the largest scales open-sourced KBs, with $34$ relation types, more than $492,600$ relations, $17$ entity types, and more than $51,000$ entities. Specifically, HiDy's knowledge is extracted from various data sources with dynamic duplications and conflict resolution techniques. For example, knowledge in \textit{Micro} branch such as partnership is mainly extracted from financial news, \textit{Macro} knowledge such as industrial strategies is extracted from government planning documents. Moreover, HiDy offers support for monthly KB updates and provides open-source extraction algorithms along with pre-trained models and a demo website.

\subsection{Open-domain KBs} Open-domain KBs are large-scale databases that encompass extensive knowledge from diverse subjects and fields. Some researchers~\cite{deng2019knowledge, feng2019temporal, sawhney2021stock, sawhney2020deep} extracted the company relationships from Wikidata~\cite{vrandevcic2014wikidata}, one of the biggest open-domain KBs. According to the claim made by Feng et al.~\cite{feng2019temporal}, Wikidata contains 58 types of first-order and second-order relations between companies in the NASDAQ and NYSE markets, such as ``Owned by”, ``Subsidiary”, ``Parent organization”, ``Founded by” and ``Follows”. The Global Knowledge Graph (GKG)\footnotemark[9] is a comprehensive network that connects individuals, organizations, locations, events, themes, and emotions from global news sources into a single massive network. Company relationships and time series of events associated with each company can be retrieved from the GKG via BigQuery~\cite{ang2022learning}.

\section{Challenges and Future Directions}\label{sec:fut}
In this section, we will discuss several open issues and future directions in knowledge-enhanced stock price prediction.

\subsection{Knowledge Acquisition}


\subsubsection{Automatical knowledge selection}
As the raw stock knowledge data contains indispensable noise, it is necessary to select the most informative and useful subset, requiring financial expertise to preliminarily discern and classify fine-grained boundaries. For instance, the methods based on high-order inter-stock correlations depend on high-confidence explicit information to categorize stocks into different group sets (i.e., hyperedges), and the methods based on exter-stock interactions ~\cite{li2019individualized, xu2021hist,zhou2022temporal} devote sufficient finance expertise to designing meta-paths. For these graph-based methods, the predefined graph substructures actually serve as prior knowledge that helps extract useful correlations and discard noise. The ongoing research requires either a strong foundation of financial expertise or great human labor in trial and error. Also, a method designed for a stock market may only work under some special conditions but cannot accommodate other markets well.

To lower the bar and the cost of knowledge acquisition, a fancy framework should automatically acquire the most informative knowledge from a complete knowledge base without being preprocessed by human-crafted structures. There are two potential ways to realize the desirable framework. AutoML~\cite{He2019AutoMLAS, Yao2018TakingHO} is one of the appealing solutions. With the knowledge structure deemed as a hyperparameter, one can turn to techniques such as meta-learning~\cite{Hospedales2020MetaLearningIN} where a meta-learner is expected to replace experts to select an optimal knowledge subset. The second idea is inspired by the recent great progress of Large Language Models (LLMs), such as ChatGPT. Similar to human learning, LLMs are fed by numerous raw data but automatically memorize the most useful knowledge through training. With the crucial stock knowledge condensed into its model parameters, it is promising for a brilliant LLM to do inference like an experienced portfolio manager and realize accurate stock prediction. Most recently, quantitative researchers have been paying attention to the applications of LLMs on stock prediction tasks and achieved considerable performance~\cite{Chen2023ChatGPTIG, LopezLira2023CanCF, Xie2023TheWS, Zhang2023UnveilingTP}. One of the future directions is to design prompts to make LLMs targeted at specific stock prediction tasks and even specific individual stocks.

\subsubsection{Imputing knowledge from multi-granularity and irregular time series}
For knowledge-enhanced stock predictions, one popular paradigm is to learn representation vectors from external knowledge as new features and then attach them to traditional technical indicators (e.g., opening prices, closing prices, and volumes). In this paradigm, we need to acquire external knowledge at all time points aligned with technical indicators. However, external knowledge data often emerge at a lower frequency than technical indicators. 
For example, in the case of daily price prediction, technical indicators are sampled on the basis of one day, while textual features from financial reports are sampled once a quarter in a coarser granularity. As for sentiment scores and event embeddings, tweets and news about a specific stock are reported irregularly and are often absent for days. Similarly, the updating frequency of event graphs cannot keep the same as the technical indicators. To impute the absent knowledge, a common practice is to fill the missing values by mean values or interpolation, transforming a coarse-grained or irregular time series into a regular one in the same granularity as stock prices. Such heuristic imputation ways can easily result in noise and require expertise to manually design.

As irregular time series have attracted research interests for decades, there are plentiful approaches to irregular time series imputation~\cite{Shukla2020ASO}, which can also be extended to financial data. However, it is still challenging and expensive to align irregular external knowledge with hourly technical indicators in the scenario of high-frequency trading. Thus, an appealing direction is to directly model the impact of tweets and events, instead of the paradigm of feature concatenation. As the Hawkes process~\cite{hawkes1971spectra} has been widely used to model event streams, one can apply Hawkes process-based methods~\cite{neural_hawkes_process} to financial events. Recently, Sawhney et al.~\cite{sawhney2021stock} applied the Hawkes process to extract temporal features from historical stock prices, which lead to significant improvement. Hou et al.~\cite{CMLF} proposed a gate mechanism to fuse the
multi-granularity technical indicators. Still, there are very few studies dealing with multi-granularity and irregular time series of external financial knowledge.

\subsection{Knowledge Incorporation}
\subsubsection{Intergrating Multimodal Knowledge}
Though existing works have explored almost all kinds of external knowledge in stock predictions, they usually focus on one of them but leave an open problem unresolved: 
\textit{whether we can combine different kinds of external knowledge for improvement?}

In practice, the effectiveness of knowledge varies by the quality of source data (e.g., the authenticity of stock news and the sparsity of stock relations). A model designed for a specific kind of knowledge may fail at some periods when the acquired knowledge is noisy and misleading. By contrast, a bunch of models that incorporate different knowledge are hard to fail simultaneously. Therefore, it is of vital significance to involve diverse external knowledge so as to analyze the stock market comprehensively. In many real-world trading firms, there are also multiple trading experts with diverse backgrounds that make predictions independently or collaboratively. Likewise, one can adopt ensemble learning over several knowledge-enhanced models, taking the combination of their outputs as the final investment decision. Alternatively, another promising direction is to adopt multimodal learning~\cite{Baltruaitis2017MultimodalML} and design a unified model that simultaneously incorporates all kinds of acquired knowledge. One can learn more informative stock representations by combining stock information from different modalities (e.g., financial texts, time-series images, and stock graphs), which provide complementary context for the same stocks and the global stock market.

\subsubsection{Generalization under Concept Drifts}
Existing works treat external knowledge as additional features of each individual stock, while we argue that external knowledge can also be taken as high-level information of the stock market which provides opportunities to tackle \textit{concept drifts}. It is well accepted that stock markets are highly non-stationary environments where the data distribution easily shifts over time~\cite{li2022ddg,LLF2021}. Such phenomena known as concept drifts cause the performance of stock prediction models to degrade along time~\cite{LLF2021}, which have attracted growing research interests in recent years. There are mainly two ideas to prevent performance degradation. The first way is to encourage the model to learn inductive bias across all periods, instead of overfitting the transductive bias and noise on the training data. To this end, AdaRNN~\cite{du2021adarnn} proposed to learn common rules shared by diverse periods while some studies~\cite{LLF2021, zhan2022meta, DoubleAdapt} adopt meta-learning to learn common patterns across past stock data and future stock data. The second appealing solution is to periodically retrain the model with new patterns acquired online. As there is still a distribution gap between the retraining data and the following test data, DDG-DA~\cite{li2022ddg} and DoubleAdapt~\cite{DoubleAdapt} proposed data adaptation methods to further mitigate the effects of distribution shifts. Nevertheless, these methods are meant to tackle predictable concept drifts.

Since unpredictable concepts drifts are still challenging to deal with, it is necessary to expand our horizon to external knowledge rather than sticking to the limited scope of traditional technical indicators. Intuitively, the new emerging knowledge, such as the changes in policies and stock relations, can reveal future tendencies of the stock market in advance of methods based on stock prices and other technical indicators. In effect, some news and changes themselves are the cause of concept drifts, which dominate the behaviors of investors. With such crucial knowledge leveraged, it is possible to make more concept drifts predictable and facilitate better model adaptation.

\subsection{Others}

\subsubsection{Explainability of knowledge-enhanced stock prediction models}
The explainability of stock price prediction is a desirable property to convince investors that the prediction is made on the basis of reliable factors (e.g., influential news or competitive corporations). This requires the quantitative investment framework to be able to demonstrate how individual stocks respond to different types of knowledge at different times, as well as their interdependence with other stocks~\cite{guo2022quant}. Deng et al.~\cite{deng2019knowledge} proposed a human-centric justification explanation method~\cite{biran2017human} to visualize the impact of events and the triples in the input graph associated with events. However, Guo et al.~\cite{guo2022quant} mentioned the explanation in the quantitive investment scenario should consider three aspects: individual stock, special time points, and factors including external knowledge. Each aspect separately indicates the correlation of the individual stock with each other, the time points of extreme market or breaking events, and the factors that can describe the individual stock and the stock market. Improving the explainability of stock price prediction can provide investors with a better understanding of the underlying factors that drive stock prices and make more informed investment decisions.

\subsubsection{Open-source platforms}

Existing knowledge-enhanced stock price prediction models often lack publicly available datasets and codes, making it challenging to compare their performance. To address this challenge, the development of open-source platforms can facilitate fair model comparisons by providing universal input data and evaluation methods. For instance, Qlib\footnotemark[12]\footnotetext[12]{https://github.com/microsoft/qlib}, an AI-oriented quantitative investment platform, offers a comprehensive pipeline encompassing data processing, model training, and backtesting, enabling users to compare diverse prediction models. However, it should be noted that Qlib lacks the incorporation of external knowledge. An alternative solution is K-Quant~\cite{Kquant}, a new knowledge-enhanced quantitative investment platform, which provides dynamic financial knowledge, stock prediction models, and interpretability methods.



\begin{acks}
This work is partially supported by the National Key Research and Development Program of China (2022YFE0200500), the National Science Foundation of China (NSFC) under Grant No. U22B2060, the Hong Kong RGC GRF Project 16213620, RIF Project R6020-19, AOE Project AoE/E-603/18, Theme-based project TRS T41-603/20R, China NSFC No. 61729201, Guangdong Basic and Applied Basic Research Foundation 2019B151530001, Hong Kong ITC ITF grants MHX/078/21 and PRP/004/22FX, Microsoft Research Asia Collaborative Research Grant, HKUST-Webank joint research lab grants, and HKUST Global Strategic Partnership Fund (2021 SJTU-HKUST).
\end{acks}

\printbibliography

@article{Zhang2023UnveilingTP,
  title={Unveiling the Potential of Sentiment: Can Large Language Models Predict Chinese Stock Price Movements{?}},
  author={Haohan Zhang and Fengrui Hua and Chengjin Xu and Jian Guo and Hao Kong and Ruiting Zuo},
  journal={arXiv:2306.14222},
  year={2023}
}

@article{Xie2023TheWS,
  title={The Wall Street Neophyte: A Zero-Shot Analysis of ChatGPT Over MultiModal Stock Movement Prediction Challenges},
  author={Qianqian Xie and Weiguang Han and Yanzhao Lai and Min Peng and Jimin Huang},
  journal={arXiv:2304.05351},
  year={2023}
}

@article{Chen2023ChatGPTIG,
  title={ChatGPT Informed Graph Neural Network for Stock Movement Prediction},
  author={Zihan Chen and Lei Zheng and Chengyu Lu and Jialu Yuan and Dixian Zhu},
  journal={arXiv:2306.03763},
  year={2023}
}

@article{lin2018domain,
  title={Domain-aware multi-truth discovery from conflicting sources},
  author={Lin, Xueling and Chen, Lei},
  journal={Proc. of the VLDB Endowment},
  volume={11},
  number={5},
  pages={635--647},
  year={2018}
}

@article{He2019AutoMLAS,
  title={AutoML: A Survey of the State-of-the-Art},
  author={Xin He and Kaiyong Zhao and Xiaowen Chu},
  journal={Knowledge-Based Systems},
  year={2021},
  volume = {212},
  pages = {106622},
}

@article{Yao2018TakingHO,
  title={Taking Human out of Learning Applications: A Survey on Automated Machine Learning},
  author={Quanming Yao and Mengshuo Wang and Hugo Jair Escalante and Isabelle M Guyon and Yi-Qi Hu and Yu-Feng Li and Wei-Wei Tu and Qiang Yang and Yang Yu},
  journal={arXiv:1810.13306},
  year={2018}
}

@article{Hospedales2020MetaLearningIN,
  author = {Timothy M. Hospedales and Antreas Antoniou and Paul Micaelli and Amos J. Storkey},
  journal = {IEEE Transactions on Pattern Analysis and Machine Intelligence},
  pages = {5149-5169},
  title = {Meta-Learning in Neural Networks: A Survey},
  year = {2022},
  volume = {44}
}

@article{LopezLira2023CanCF,
  title={Can ChatGPT Forecast Stock Price Movements? Return Predictability and Large Language Models},
  author={Alejandro Lopez-Lira and Yuehua Tang},
  journal={arXiv:2304.07619},
  year={2023}
}

@inproceedings{lil2022flag,
  title={FLAG: Stock Movement Prediction via Fusing Logic and Semantic Graphs of Financial News},
  author={Lil, Angela and Liu, Jiduan and Lil, Yuyong and Meng, Fan},
  booktitle={2022 IEEE Intl. Conf. on Data Mining Workshops},
  pages={190--197},
  year={2022},
  organization={IEEE}
}

@InProceedings{LLF2021,
  author    = {Xiaoyu You and Mi Zhang and Daizong Ding and Fuli Feng and Yuanmin Huang},
  booktitle = {Proc. of the 30th {ACM} Intl. Conf. on Information {\&} Knowledge Management {(CIKM'21)}},
  title     = {Learning to Learn the Future: Modeling Concept Drift in Time Series Prediction},
  pages = {2434–2443},
  year      = {2021}
}

@InProceedings{DoubleAdapt,
  author       = {Lifan Zhao and Shuming Kong and Yanyan Shen},
  booktitle    = {Proc. of the 29th {ACM} {SIGKDD} Conf. on Knowledge Discovery and Data Mining},
  title        = {{DoubleAdapt}: A Meta-learning Approach to Incremental Learning for Stock Trend Forecasting},
  year         = {2023}
}

@inproceedings{sil2013re,
  title={Re-ranking for joint named-entity recognition and linking},
  author={Sil, Avirup and Yates, Alexander},
  booktitle={Proc. of the 22nd ACM Intl. Conf. on Information {\&} Knowledge Management {(CIKM'13)}},
  pages={2369--2374},
  year={2013}
}

@inproceedings{han-etal-2019-opennre,
    title = {{O}pen{NRE}: An Open and Extensible Toolkit for Neural Relation Extraction},
    author = {Han, Xu  and
      Gao, Tianyu  and
      Yao, Yuan  and
      Ye, Deming  and
      Liu, Zhiyuan  and
      Sun, Maosong},
    booktitle = {Proc. of the EMNLP-IJCNLP},
    year = {2019},
    pages = {169--174}
}

@article{Shukla2020ASO,
  title={A Survey on Principles, Models and Methods for Learning from Irregularly Sampled Time Series: From Discretization to Attention and Invariance},
  author={Satya Narayan Shukla and Benjamin M Marlin},
  journal={arXiv:2012.00168},
  year={2020}
}

@inproceedings{CMLF,
    author = {Hou, Min and Xu, Chang and Liu, Yang and Liu, Weiqing and Bian, Jiang and Wu, Le and Li, Zhi and Chen, Enhong and Liu, Tie-Yan},
    title = {Stock Trend Prediction with Multi-Granularity Data: A Contrastive Learning Approach with Adaptive Fusion},
    year = {2021},
    booktitle = {Proc. of the 30th ACM Intl. Conf. on Information {\&} Knowledge Management {(CIKM'21)}},
    pages = {700–709},
    numpages = {10}
}

@article{hawkes1971spectra,
  title={Spectra of some self-exciting and mutually exciting point processes},
  author={Hawkes, Alan G},
  journal={Biometrika},
  volume={58},
  number={1},
  pages={83--90},
  year={1971}
}

@inproceedings{neural_hawkes_process,
author = {Mei, Hongyuan and Eisner, Jason},
title = {The Neural Hawkes Process: A Neurally Self-Modulating Multivariate Point Process},
year = {2017},
booktitle = {Proc. of the 31st Intl. Conf. on Neural Information Processing Systems {(NIPS'17)}},
pages = {6757–6767}
}

@article{li2020multimodal,
  author={Li, Qing and Tan, Jinghua and Wang, Jun and Chen, Hsinchun},
  journal={IEEE Transactions on Knowledge and Data Engineering},
  title={A Multimodal Event-Driven LSTM Model for Stock Prediction Using Online News}, 
  year={2021},
  volume={33},
  number={10},
  pages={3323-3337}}

@inbook{yadati2019hypergcn,
    author = {Yadati, Naganand and Nimishakavi, Madhav and Yadav, Prateek and Nitin, Vikram and Louis, Anand and Talukdar, Partha},
    title = {HyperGCN: A New Method of Training Graph Convolutional Networks on Hypergraphs},
    year = {2019},
    booktitle = {Proc. of the 33rd Intl. Conf. on Neural Information Processing Systems {(NIPS'19)}},
    volume={32},
    numpages = {12}
}

@INPROCEEDINGS{Li2022hyper,
  author={Li, Xiaojie and Cui, Chaoran and Cao, Donglin and Du, Juan and Zhang, Chunyun},
  booktitle={ICASSP 2022 IEEE Intl. Conf. on Acoustics, Speech and Signal Processing}, 
  title={Hypergraph-Based Reinforcement Learning for Stock Portfolio Selection}, 
  year={2022},
  pages={4028-4032}}

@article{zhao2022stock,
  title={Stock Movement Prediction Based on Bi-Typed Hybrid-Relational Market Knowledge Graph via Dual Attention Networks},
  author={Yu Zhao and Huaming Du and Ying Liu and Shaopeng Wei and Xingyan Chen and Fuzhen Zhuang and Qing Li and Ji Liu and Gang Kou},
  journal={IEEE Transactions on Knowledge and Data Engineering},
  year={2022},
  volume={35},
  pages={8559-8571}
}

@article{ma2023multi,
  title={Multi-source aggregated classification for stock price movement prediction},
  author={Ma, Yu and Mao, Rui and Lin, Qika and Wu, Peng and Cambria, Erik},
  journal={Information Fusion},
  volume={91},
  pages={515--528},
  year={2023}
}

@article{Baltruaitis2017MultimodalML,
  title={Multimodal Machine Learning: A Survey and Taxonomy},
  author={Tadas Baltru{\v{s}}aitis and Chaitanya Ahuja and Louis-Philippe Morency},
  journal={IEEE Transactions on Pattern Analysis and Machine Intelligence},
  year={2019},
  volume={41},
  pages={423-443}
}

@article{de2009predicting,
  title={Predicting the Brazilian stock market through neural networks and adaptive exponential smoothing methods},
  author={De Faria, EL and Albuquerque, Marcelo P and Gonzalez, JL and Cavalcante, JTP and Albuquerque, Marcio P},
  journal={Expert Systems with Applications},
  volume={36},
  number={10},
  pages={12506--12509},
  year={2009}
}

@article{lu2020cnn,
author = {Lu, Wenjie and Li, Jiazheng and Li, Yifan and Sun, Aijun and Wang, Jingyang and Hassanien, Abd {E. I.-B}aset},
title = {A CNN-LSTM-Based Model to Forecast Stock Prices},
year = {2020},
volume = {2020},
journal = {Complexity}, 
month = {jan}, 
numpages = {10}
}

@inproceedings{qin2019you,
    title = "What You Say and How You Say It Matters: Predicting Stock Volatility Using Verbal and Vocal Cues",
    author = "Qin, Yu  and
      Yang, Yi",
    booktitle = "Proc. of the 57th Annual Meeting of the Association for Computational Linguistics",
    year = "2019",
    pages = "390--401"
}

@article{vaiz2016study,
  title={A study on technical indicators in stock price movement prediction using decision tree algorithms},
  author={Vaiz, J Sharmila and Ramaswami, M},
  journal={American Journal of Engineering Research (AJER)},
  volume={5},
  number={12},
  pages={207--212},
  year={2016}
}

@article{fenghua2014stock,
  title={Stock price prediction based on SSA and SVM},
  author={Fenghua, WEN and Jihong, XIAO and Zhifang, HE and Xu, GONG},
  journal={Procedia Computer Science},
  volume={31},
  pages={625--631},
  year={2014}
}

@inproceedings{ariyo2014stock,
  title={Stock price prediction using the ARIMA model},
  author={Ariyo, Adebiyi A and Adewumi, Adewumi O and Ayo, Charles K},
  booktitle={2014 UKSim-AMSS 16th Intl. Conf. on Computer Modelling and Simulation},
  pages={106--112},
  year={2014}
}

@misc{Kquant,
  title = {K-{Q}uant: A Platform of Temporal Financial Knowledge-enhanced Quantitative Investment},
  author = {K-{Q}uant},
  published = {{\url{https://github.com/K-Quant}}},
  year = {2023},
}

@article{li2018stock,
  title={Stock prediction via sentimental transfer learning},
  author={Li, Xiaodong and Xie, Haoran and Lau, Raymond YK and Wong, Tak-Lam and Wang, Fu-Lee},
  journal={IEEE Access},
  volume={6},
  pages={73110--73118},
  year={2018},
}

@article{hagenau2013automated,
  title={Automated news reading: Stock price prediction based on financial news using context-capturing features},
  author={Hagenau, Michael and Liebmann, Michael and Neumann, Dirk},
  journal={Decision Support Systems},
  volume={55},
  number={3},
  pages={685--697},
  year={2013}
}

@article{luss2015predicting,
  title={Predicting abnormal returns from news using text classification},
  author={Luss, Ronny and {d'{A}}spremont, Alexandre},
  journal={Quantitative Finance},
  volume={15},
  number={6},
  pages={999--1012},
  year={2015}
}

@article{feng2019temporal, 
author = {Feng, Fuli and He, Xiangnan and Wang, Xiang and Luo, Cheng and Liu, Yiqun and Chua, Tat-Seng}, 
title = {Temporal Relational Ranking for Stock Prediction}, 
year = {2019}, 
volume = {37}, 
number = {2}, 
pages={1--30},
journal = {ACM Transactions on Information Systems}
}

@inproceedings{yang2020html,
  title={Html: Hierarchical transformer-based multi-task learning for volatility prediction},
  author={Yang, Linyi and Ng, Tin Lok James and Smyth, Barry and Dong, Riuhai},
  booktitle={Proc. of The Web Conf. 2020},
  pages={441--451},
  year={2020}
}

@article{guo2022quant,
  title={Quant 4.0: Engineering Quantitative Investment with Automated, Explainable and Knowledge-driven Artificial Intelligence},
  author={Guo, Jian and Wang, Saizhuo and Ni, Lionel M and Shum, Heung-Yeung},
  journal={arXiv:2301.04020},
  year={2022}
}

@article{skarding2021foundations,
  title={Foundations and modeling of dynamic networks using dynamic graph neural networks: A survey},
  author={Skarding, Joakim and Gabrys, Bogdan and Musial, Katarzyna},
  journal={IEEE Access},
  volume={9},
  pages={79143--79168},
  year={2021}
}

@inproceedings{xiang2022temporal,
  title={Temporal and Heterogeneous Graph Neural Network for Financial Time Series Prediction},
  author={Xiang, Sheng and Cheng, Dawei and Shang, Chencheng and Zhang, Ying and Liang, Yuqi},
  booktitle={Proc. of the 31st {ACM} Intl. Conf. on Information Knowledge Management {(CIKM'22)}},
  pages={3584--3593},
  year={2022}
}

@inproceedings{biran2017human,
  title={Human-Centric Justification of Machine Learning Predictions},
  author={Biran, Or and McKeown, Kathleen R},
  booktitle={Proc. of the 26th Intl. Joint Conf. on Artificial Intelligence {(IJCAI'17)}},
  volume={2017},
  pages={1461--1467},
  year={2017}
}

@article{fama1970efficient,
  title={Efficient capital markets: A review of theory and empirical work},
  author={Fama, Eugene F},
  journal={The Journal of Finance},
  volume={25},
  number={2},
  pages={383--417},
  year={1970}
}

@inproceedings{zhu2023hit,
  title={HIT: An Effective Approach to Build a Dynamic Financial Knowledge Base},
  author={Zhu, Xinyi and Xin, Hao and Shen, Yanyan and Chen, Lei},
  booktitle={DASFAA},
  pages={716--731},
  year={2023}
}

@inproceedings{chen2018incorporating,
  title={Incorporating corporation relationship via graph convolutional neural networks for stock price prediction},
  author={Chen, Yingmei and Wei, Zhongyu and Huang, Xuanjing},
  booktitle={Proc. of the 27th ACM Intl. Conf. on Information and Knowledge Management {(CIKM'18)}},
  pages={1655--1658},
  year={2018}
}

@article{xu2021hist,
  title={HIST: A Graph-based Framework for Stock Trend Forecasting via Mining Concept-Oriented Shared Information},
  author={Xu, Wentao and Liu, Weiqing and Wang, Lewen and Xia, Yingce and Bian, Jiang and Yin, Jian and Liu, Tie-Yan},
  journal={arXiv:2110.13716},
  year={2021}
}

@article{saha2021stock,
  title={Stock ranking prediction using list-wise approach and node embedding technique},
  author={Saha, Suman and Gao, Junbin and Gerlach, Richard},
  journal={IEEE Access},
  volume={9},
  pages={88981--88996},
  year={2021},
  publisher={IEEE}
}

@inproceedings{cheng2020knowledge,
  title={Knowledge graph-based event embedding framework for financial quantitative investments},
  author={Cheng, Dawei and Yang, Fangzhou and Wang, Xiaoyang and Zhang, Ying and Zhang, Liqing},
  booktitle={Proc. of the 43rd Intl. ACM SIGIR Conf. on Research and Development in Information Retrieval {(SIGIR'20)}},
  pages={2221--2230},
  year={2020}
}

@article{sarmah2022learning,
  title={Learning Embedded Representation of the Stock Correlation Matrix using Graph Machine Learning},
  author={Sarmah, Bhaskarjit and Nair, Nayana and Mehta, Dhagash and Pasquali, Stefano},
  journal={arXiv{:}2207.07183},
  year={2022}
}

@inproceedings{sawhney2020deep,
  title={Deep attentive learning for stock movement prediction from social media text and company correlations},
  author={Sawhney, Ramit and Agarwal, Shivam and Wadhwa, Arnav and Shah, Rajiv},
  booktitle={Proc. of the 2020 Conf. on Empirical Methods in Natural Language Processing {(EMNLP'20)}},
  pages={8415--8426},
  year={2020}
}

@inproceedings{perozzi2014deepwalk,
  title={Deepwalk: Online learning of social representations},
  author={Perozzi, Bryan and Al-Rfou, Rami and Skiena, Steven},
  booktitle={Proc. of the 20th ACM SIGKDD Intl. Conf. on Knowledge Discovery and Data Mining {(KDD'14)}},
  pages={701--710},
  year={2014}
}

@inproceedings{grover2016node2vec,
  title={node2vec: Scalable feature learning for networks},
  author={Grover, Aditya and Leskovec, Jure},
  booktitle={Proc. of the 22nd ACM SIGKDD Intl. Conf. on Knowledge Discovery and Data Mining {(KDD'16)}},
  pages={855--864},
  year={2016}
}

@inproceedings{tang2015line,
  title={Line: Large-scale information network embedding},
  author={Tang, Jian and Qu, Meng and Wang, Mingzhe and Zhang, Ming and Yan, Jun and Mei, Qiaozhu},
  booktitle={Proc. of the 24th Intl. Conf. on World Wide Web {(WWW'15)}},
  pages={1067--1077},
  year={2015}
}

@inproceedings{mikolov2013efficient,
  title={Efficient Estimation of Word Representations in Vector Space},
  author={Mikolov, Tomas and Chen, Kai and Corrado, Greg and Dean, Jeffrey},
  booktitle={Intl. Conf. on Learning Representations},
  year={2013}
}

@article{hu2021survey,
  title={A survey of forex and stock price prediction using deep learning},
  author={Hu, Zexin and Zhao, Yiqi and Khushi, Matloob},
  journal={Applied System Innovation},
  volume={4},
  number={1},
  pages={9},
  year={2021}
}

@inproceedings{hossain2018hybrid,
  title={Hybrid deep learning model for stock price prediction},
  author={Hossain, Mohammad Asiful and Karim, Rezaul and Thulasiram, Ruppa and Bruce, Neil DB and Wang, Yang},
  booktitle={2018 IEEE Symposium Series on Computational Intelligence (SSCI)},
  pages={1837--1844},
  year={2018}
}

@article{wang2021review,
  title={A review on graph neural network methods in financial applications},
  author={Wang, Jianian and Zhang, Sheng and Xiao, Yanghua and Song, Rui},
  journal={Journal of Data Science},
  volume = {20},
  number = {2},
  year = {2022}
}

@inproceedings{akita2016deep,
  title={Deep learning for stock prediction using numerical and textual information},
  author={Akita, Ryo and Yoshihara, Akira and Matsubara, Takashi and Uehara, Kuniaki},
  booktitle={2016 {IEEE/ACIS} 15th Intl. Conf. on Computer and Information Science (ICIS)},
  pages={1--6},
  year={2016},
}

@inproceedings{yang2015network,
  title={Network representation learning with rich text information},
  author={Yang, Cheng and Liu, Zhiyuan and Zhao, Deli and Sun, Maosong and Chang, Edward},
  booktitle={Proc. of the 24th Intl. Conf. on Artificial Intelligence {(IJCAI'15)}},
  year={2015},
  pages = {2111–2117}
}

@inproceedings{li2021modeling,
  title={Modeling the stock relation with graph network for overnight stock movement prediction},
  author={Li, Wei and Bao, Ruihan and Harimoto, Keiko and Chen, Deli and Xu, Jingjing and Su, Qi},
  booktitle={Proc. of the 29th Intl. Joint Conference on Artificial Intelligence {(IJCAI'20)}},
  pages={4541--4547},
  year={2020}
}

@inproceedings{pan2016tri,
  title={Tri-party deep network representation},
  author={Pan, Shirui and Wu, Jia and Zhu, Xingquan and Zhang, Chengqi and Wang, Yang},
  journal={Proc. of the 25th Intl. Joint Conf. on Artificial Intelligence {(IJCAI'16)}},
  pages = {1895–1901},
  year={2016}
}

@inproceedings{ganea2017deep,
    title = "Deep Joint Entity Disambiguation with Local Neural Attention",
    author = "Ganea, Octavian-Eugen  and Hofmann, Thomas",
    booktitle = "Proc. of the 2017 Conf. on Empirical Methods in Natural Language Processing",
    year = "2017",
    pages = "2619--2629",
}

@article{pozzi2013spread,
  title={Spread of risk across financial markets: better to invest in the peripheries},
  author={Pozzi, Francesco and Di Matteo, Tiziana and Aste, Tomaso},
  journal={Scientific reports},
  volume={3},
  number={1},
  pages={1665},
  year={2013}
}

@article{aste2010correlation,
  title={Correlation structure and dynamics in volatile markets},
  author={Aste, Tomaso and Shaw, William and Di Matteo, Tiziana},
  journal={New Journal of Physics},
  volume={12},
  number={8},
  pages={085009},
  year={2010}
}

@article{chi2010network,
  title={A network perspective of the stock market},
  author={Chi, K Tse and Liu, Jing and Lau, Francis CM},
  journal={Journal of Empirical Finance},
  volume={17},
  number={4},
  pages={659-667},
  year={2010}
}

@inproceedings{wang2021global,
  title={Global Semantics with Boundary Constraint Knowledge Graph for Chinese Financial Event Detection},
  author={Wang, Yin and Xia, Nan and Luo, Xiangfeng and Li, Jinhui},
  booktitle={2021 IEEE Intl. Conf. on Big Knowledge},
  pages={281-289},
  year={2021}
}

@article{vrandevcic2014wikidata,
  title={Wikidata: a free collaborative knowledgebase},
  author={Vrande{\v{c}}i{\'{c}}, Denny and Kr{\'{o}}tzsch, Markus},
  journal={Communications of the ACM},
  volume={57},
  number={10},
  pages={78--85},
  year={2014}
}

@inproceedings{bollacker2008freebase,
  title={Freebase: a collaboratively created graph database for structuring human knowledge},
  author={Bollacker, Kurt and Evans, Colin and Paritosh, Praveen and Sturge, Tim and Taylor, Jamie},
  booktitle={Proc. of the 2008 ACM SIGMOD Intl. Conf. on Management of Data {(SIGMOD'08)}},
  pages={1247--1250},
  year={2008}
}

@inproceedings{deng2019knowledge,
  title={Knowledge-driven stock trend prediction and explanation via temporal convolutional network},
  author={Deng, Shumin and Zhang, Ningyu and Zhang, Wen and Chen, Jiaoyan and Pan, Jeff Z and Chen, Huajun},
  booktitle={Companion Proc. of The 2019 World Wide Web Conf. {(WWW'19)}},
  pages={678--685},
  year={2019}
}

@article{tumminello2005tool,
  title={A tool for filtering information in complex systems},
  author={Tumminello, Michele and Aste, Tomaso and Di Matteo, Tiziana and Mantegna, Rosario N},
  journal={Proc. of the National Academy of Sciences},
  volume={102},
  number={30},
  pages={10421--10426},
  year={2005}
}

@article{zhu2009technical,
  title={Technical analysis: An asset allocation perspective on the use of moving averages},
  author={Zhu, Yingzi and Zhou, Guofu},
  journal={Journal of Financial Economics},
  volume={92},
  number={3},
  pages={519--544},
  year={2009}
}

@inproceedings{li2019individualized,
  title={Individualized indicator for all: Stock-wise technical indicator optimization with stock embedding},
  author={Li, Zhige and Yang, Derek and Zhao, Li and Bian, Jiang and Qin, Tao and Liu, Tie-Yan},
  booktitle={Proc. of the 25th ACM SIGKDD Intl. Conf. on Knowledge Discovery {\&} Data Mining {(SIGKDD'19)}},
  pages={894--902},
  year={2019}
}

@inproceedings{xu2021rest,
  title={Rest: Relational event-driven stock trend forecasting},
  author={Xu, Wentao and Liu, Weiqing and Xu, Chang and Bian, Jiang and Yin, Jian and Liu, Tie-Yan},
  booktitle={Proc. of the Web Conf. 2021 {(WWW'21)}},
  pages={1--10},
  year={2021}
}

@inproceedings{cer2018universal,
  title={Universal sentence encoder for English},
  author={Cer, Daniel and Yang, Yinfei and Kong, Sheng-yi and Hua, Nan and Limtiaco, Nicole and John, Rhomni St and Constant, Noah and Guajardo-Cespedes, Mario and Yuan, Steve and Tar, Chris and others},
  booktitle={Proc. of the 2018 Conf. on Empirical Methods in Natural Language Processing: System Demonstrations {(EMNLP'18)}},
  pages={169--174},
  year={2018}
}

@article{zhou2020graph,
  title={Graph neural networks: A review of methods and applications},
  author={Zhou, Jie and Cui, Ganqu and Hu, Shengding and Zhang, Zhengyan and Yang, Cheng and Liu, Zhiyuan and Wang, Lifeng and Li, Changcheng and Sun, Maosong},
  journal={AI open},
  volume={1},
  pages={57--81},
  year={2020}
}

@inproceedings{kipf2016semi,
  author = {Kipf, Thomas N and Welling, Max},
  title  = {Semi-Supervised Classification with Graph Convolutional Networks},
  booktitle = {5th Intl. Conf. on Learning Representations {(ICLR'17)}},
  year = {2017}
}

@article{
  velivckovic2017graph,
  title={Graph Attention Networks},
  author={Veli{\v{c}}kovi{\'{c}}, Petar and Cucurull, Guillem and Casanova, Arantxa and Romero, Adriana and Li{\'{o}}, Pietro and Bengio, Yoshua},
  journal={Intl. Conf. on Learning Representations {(ICLR'18)}},
  year={2018}
}

@article{matsunaga2019exploring,
  title={Exploring graph neural networks for stock market predictions with rolling window analysis},
  author={Matsunaga, Daiki and Suzumura, Toyotaro and Takahashi, Toshihiro},
  journal={arXiv{:}1909.10660},
  year={2019}
}

@article{yan2020development,
  title={Development of stock networks using part mutual information and australian stock market data},
  author={Yan, Yan and Wu, Boyao and Tian, Tianhai and Zhang, Hu},
  journal={Entropy},
  volume={22},
  number={7},
  pages={773},
  year={2020},
}

@INPROCEEDINGS{9670610,
  author={Xie, Chunfu},
  booktitle={2021 Intl. Conf. on Digital Society and Intelligent Systems}, 
  title={Stock Prediction Based On Event Graph}, 
  year={2021},
  volume={},
  number={},
  pages={240-244}
 }

@article{li2019portfolio,
  title={Portfolio optimization based on network topology},
  author={Li, Yan and Jiang, Xiong-Fei and Tian, Yue and Li, Sai-Ping and Zheng, Bo},
  journal={Physica A: Statistical Mechanics and its Applications},
  volume={515},
  pages={671--681},
  year={2019}
}

@article{sim2019deep,
  title={Is deep learning for image recognition applicable to stock market prediction{?}},
  author={Sim, Hyun Sik and Kim, Hae In and Ahn, Jae Joon},
  journal={Complexity},
  volume={2019},
  year={2019}
}

@article{raffinot2017hierarchical,
  title={Hierarchical clustering-based asset allocation},
  author={Raffinot, Thomas},
  journal={The Journal of Portfolio Management},
  volume={44},
  number={2},
  pages={89-99},
  year={2017}
}

@inproceedings{ding2016knowledge,
  title={Knowledge-driven event embedding for stock prediction},
  author={Ding, Xiao and Zhang, Yue and Liu, Ting and Duan, Junwen},
  booktitle={Proc. of {COLING} 2016, the 26th Intl. Conf. on Computational Linguistics: Technical Papers},
  pages={2133--2142},
  year={2016}
}

@article{chen2018leveraging,
  title={Leveraging social media news to predict stock index movement using RNN-boost},
  author={Chen, Weiling and Yeo, Chai Kiat and Lau, Chiew Tong and Lee, Bu Sung},
  journal={Data {\&} Knowledge Engineering},
  volume={118},
  pages={14--24},
  year={2018}
}

@inproceedings{xu2019stock,
  title={Stock prediction using deep learning and sentiment analysis},
  author={Xu, Yichuan and Keselj, Vlado},
  booktitle={2019 IEEE Intl. Conf. on Big Data},
  pages={5573--5580},
  year={2019}
}

@inproceedings{hu2018listening,
  title={Listening to chaotic whispers: A deep learning framework for news-oriented stock trend prediction},
  author={Hu, Ziniu and Liu, Weiqing and Bian, Jiang and Liu, Xuanzhe and Liu, Tie-Yan},
  booktitle={Proc. of the Eleventh ACM Intl. Conf. on Web Search and Data Mining},
  pages={261--269},
  year={2018}
}

@inproceedings{ying2020time,
  title={Time-aware graph relational attention network for stock recommendation},
  author={Ying, Xiaoting and Xu, Cong and Gao, Jianliang and Wang, Jianxin and Li, Zhao},
  booktitle={Proc. of the 29th ACM Intl. Conf. on Information {\&} Knowledge Management {(CIKM'20)}},
  pages={2281--2284},
  year={2020}
}

@inproceedings{ding2015deep,
  title={Deep learning for event-driven stock prediction},
  author={Ding, Xiao and Zhang, Yue and Liu, Ting and Duan, Junwen},
  booktitle={Proc. of the 24th Intl. Conf. on Artificial Intelligence {(IJCAI'15)}},
  pages = {2327–2333},
  year={2015}
}

@article{creamer2009link,
  title={A link mining algorithm for earnings forecast and trading},
  author={Creamer, Germ{\'{a}}n and Stolfo, Sal},
  journal={Data Mining and Knowledge Discovery},
  volume={18},
  number={3},
  pages={419--445},
  year={2009}
}

@article{lu2018herding,
  title={Herding boosts too-connected-to-fail risk in stock market of China},
  author={Lu, Shan and Zhao, Jichang and Wang, Huiwen and Ren, Ruoen},
  journal={Physica A: Statistical Mechanics and its Applications},
  volume={505},
  pages={945--964},
  year={2018}
}

@inproceedings{bisarya2022stock,
  title={Stock Price Prediction Using Corporation Network and LSTM},
  author={Bisarya, Udbhav and Parekh, Vishwas and Bhattacharjee, Shrutilipi},
  booktitle={2022 2nd Intl. Conf. on Intelligent Technologies},
  pages={1--6},
  year={2022}
}

@inproceedings{cardoso2022learning,
  title={Learning Bipartite Graphs: Heavy Tails and Multiple Components},
  author={Cardoso, Jos{\'{e}} Vin{\'{i}}cius De Miranda and Ying, Jiaxi and Palomar, Daniel P},
  booktitle={NeurIPS},
  pages = {14044--14057},
  volume = {35},
  year={2022}
}

@article{jiang2021applications,
  title={Applications of deep learning in stock market prediction: recent progress},
  author={Jiang, Weiwei},
  journal={Expert Systems with Applications},
  volume={184},
  pages={115537},
  year={2021}
}

@inproceedings{zhou2022temporal,
  title={Temporal Bipartite Graph Neural Networks for Bond Prediction},
  author={Zhou, Dan and Uddin, Ajim and Tao, Xinyuan and Shang, Zuofeng and Yu, Dantong},
  booktitle={Proc. of the Third ACM Intl. Conf. on AI in Finance},
  pages={308--316},
  year={2022}
}

@inproceedings{wilson2005opinionfinder,
  title={OpinionFinder: A system for subjectivity analysis},
  author={Wilson, Theresa and Hoffmann, Paul and Somasundaran, Swapna and Kessler, Jason and Wiebe, Janyce and Choi, Yejin and Cardie, Claire and Riloff, Ellen and Patwardhan, Siddharth},
  booktitle={Proc. of {HLT/EMNLP} 2005 interactive demonstrations},
  pages={34--35},
  year={2005}
}

@article{araci2019finbert,
  title={Finbert: Financial sentiment analysis with pre-trained language models},
  author={Araci, Dogu},
  journal={arXiv:1908.10063},
  year={2019}
}

@article{tangjitprom2011macroeconomic,
  title={Macroeconomic factors of emerging stock market: the evidence from Thailand},
  author={Tangjitprom, Nopphon},
  journal={Intl. Journal of Financial Research},
  volume={3},
  number={2},
  pages={105--114},
  year={2011}
}

@article{ogundunmade2022stock,
  title={Stock price forecasting: Machine learning models with K-fold and repeated cross validation approaches},
  author={Ogundunmade, TP and Adepoju, AA and Allam, A},
  journal={Modern Economy and Management},
  volume={1},
  number={1},
  pages={2},
  year={2022}
}

@article{chung2018genetic,
  title={Genetic algorithm-optimized long short-term memory network for stock market prediction},
  author={Chung, Hyejung and Shin, Kyung-shik},
  journal={Sustainability},
  volume={10},
  number={10},
  pages={3765},
  year={2018}
}

@inproceedings{zhang2018new,
  title={A new combined CNN-RNN model for sector stock price analysis},
  author={Zhang, Ruixun and Yuan, Zhaozheng and Shao, Xiuli},
  booktitle={2018 IEEE 42nd Annual Computer Software and Applications Conference},
  volume={2},
  pages={546--551},
  year={2018}
}

@article{chatzis2018forecasting,
  title={Forecasting stock market crisis events using deep and statistical machine learning techniques},
  author={Chatzis, Sotirios P and Siakoulis, Vassilis and Petropoulos, Anastasios and Stavroulakis, Evangelos and Vlachogiannakis, Nikos},
  journal={Expert Systems with Applications},
  volume={112},
  pages={353--371},
  year={2018}
}

@article{nikou2019stock,
  title={Stock price prediction using DEEP learning algorithm and its comparison with machine learning algorithms},
  author={Nikou, Mahla and Mansourfar, Gholamreza and Bagherzadeh, Jamshid},
  journal={Intelligent Systems in Accounting, Finance and Management},
  volume={26},
  number={4},
  pages={164--174},
  year={2019},
}

@article{chen2022gated,
  title={Gated three-tower transformer for text-driven stock market prediction},
  author={Chen, Jia and Chen, Tao and Shen, Mengqi and Shi, Yunhai and Wang, Dongjing and Zhang, Xin},
  journal={Multimedia Tools and Applications},
  pages={30093–30119},
  volume = {81},
  number = {21},
  year={2022}
}

@article{khan2020stock,
  title={Stock market prediction using machine learning classifiers and social media, news},
  author={Khan, Wasiat and Ghazanfar, Mustansar Ali and Azam, Muhammad Awais and Karami, Amin and Alyoubi, Khaled H and Alfakeeh, Ahmed S},
  journal={Journal of Ambient Intelligence and Humanized Computing},
  pages={3433--3456},
  volume = {13},
  year={2022}
}

@article{blei2003latent,
  title={Latent dirichlet allocation},
  author={Blei, David M and Ng, Andrew Y and Jordan, Michael I},
  journal={Journal of machine Learning research},
  volume={3},
  pages={993--1022},
  year={2003}
}

@article{wang2021data,
  title={Data set and evaluation of automated construction of financial knowledge graph},
  author={Wang, Wenguang and Xu, Yonglin and Du, Chunhui and Chen, Yunwen and Wang, Yijie and Wen, Hui},
  journal={Data Intelligence},
  volume={3},
  number={3},
  pages={418--443},
  year={2021}
}

@inproceedings{guo2020knowledge,
  title={Knowledge graph enhanced event extraction in financial documents},
  author={Guo, Kaihao and Jiang, Tianpei and Zhang, Haipeng},
  booktitle={2020 IEEE Intl. Conf. on Big Data},
  pages={1322--1329},
  year={2020}
}

@inproceedings{wu2020creating,
  title={Creating a large-scale financial news corpus for relation extraction},
  author={Wu, Haoyu and Lei, Qing and Zhang, Xinyue and Luo, Zhengqian},
  booktitle={2020 3rd Intl. Conf. on Artificial Intelligence and Big Data},
  pages={259--263},
  year={2020}
}

@inproceedings{jabbari2020french,
  title={A French corpus and annotation schema for named entity recognition and relation extraction of financial news},
  author={Jabbari, Ali and Sauvage, Olivier and Zeine, Hamada and Chergui, Hamza},
  booktitle={Proc. of the Twelfth Language Resources and Evaluation Conf.},
  pages={2293--2299},
  year={2020}
}

@inproceedings{sharma2022finred,
  title={Fin{RED}: A Dataset for Relation Extraction in Financial Domain},
  author={Sharma, Soumya and Nayak, Tapas and Bose, Arusarka and Meena, Ajay Kumar and Dasgupta, Koustuv and Ganguly, Niloy and Goyal, Pawan},
  booktitle={Companion Proc. of the Web Conference 2022 {(WWW'22)}},
  pages={595--597},
  year={2022}
}

@inproceedings{liu2019transformer,
  title={Transformer-based capsule network for stock movement prediction},
  author={Liu, Jintao and Lin, Hongfei and Liu, Xikai and Xu, Bo and Ren, Yuqi and Diao, Yufeng and Yang, Liang},
  booktitle={Proc. of the First Workshop on Financial Technology and Natural Language Processing},
  pages={66--73},
  year={2019}
}

@inproceedings{lee2017predict,
  title={Predict stock price with financial news based on recurrent convolutional neural networks},
  author={Lee, Che-Yu and Soo, Von-Wun},
  booktitle={2017 Conf. on Technologies and Applications of Artificial Intelligence},
  pages={160--165},
  year={2017},
}

@inproceedings{Steven2006nltk,
    author = {Bird, Steven},
    title = {NLTK: The Natural Language Toolkit},
    year = {2006},
    booktitle = {Proc. of the {COLING/ACL} on Interactive Presentation Sessions},
    pages = {69–72}
}

@article{huang2022news,
  title={News-driven stock prediction via noisy equity state representation},
  author={Huang, Heyan and Liu, Xiao and Zhang, Yue and Feng, Chong},
  journal={Neurocomputing},
  volume={470},
  pages={66--75},
  year={2022},
  publisher={Elsevier}
}

@article{kilimci2020efficient,
  title={An efficient word embedding and deep learning based model to forecast the direction of stock exchange market using twitter and financial news sites: a case of istanbul stock exchange (bist 100)},
  author={Kilimci, Zeynep Hilal and Duvar, Ramazan},
  journal={IEEE Access},
  volume={8},
  pages={188186--188198},
  year={2020}
}

@inproceedings{han2022duee,
  title={DuEE-Fin: A Large-Scale Dataset for Document-Level Event Extraction},
  author={Han, Cuiyun and Zhang, Jinchuan and Li, Xinyu and Xu, Guojin and Peng, Weihua and Zeng, Zengfeng},
  booktitle={Natural Language Processing and Chinese Computing: 11th CCF Intl. Conf.},
  pages={172--183},
  year={2022}
}

@article{gupta2014clustering,
  title={Clustering-Classification based prediction of stock market future prediction},
  author={Gupta, Abhishek and Sharma, Samidha D},
  journal={Intl. Journal of Computer Science and Information Technologies},
  volume={5},
  number={3},
  pages={2806--2809},
  year={2014}
}

@inproceedings{zheng2019doc2edag,
  title={Doc2EDAG: An end-to-end document-level framework for Chinese financial event extraction},
  author={Zheng, Shun and Cao, Wei and Xu, Wei and Bian, Jiang},
  booktitle = "Proc. of the 2019 Conf. on Empirical Methods in Natural Language Processing",
  pages = "337--346",
  year={2019}
}

@inproceedings{yang2018dcfee,
  title={Dcfee: A document-level chinese financial event extraction system based on automatically labeled training data},
  author={Yang, Hang and Chen, Yubo and Liu, Kang and Xiao, Yang and Zhao, Jun},
  booktitle = "Proc. of {ACL} 2018, System Demonstrations",
  pages={50--55},
  year={2018}
}

@inproceedings{zhao2019uer,
    title = {{UER}: An Open-Source Toolkit for Pre-training Models},
    author = {Zhao, Zhe and Chen, Hui and Zhang, Jinbin and Zhao, Xin and Liu, Tao and Lu, Wei and Chen, Xi and Deng, Haotang and Ju, Qi and Du, Xiaoyong},
    booktitle = {Proc. of the 2019 Conf. on Empirical Methods in Natural Language Processing and the 9th Intl. Joint Conf. on Natural Language Processing: System Demonstrations},
    year = {2019},
    pages = {241--246}
}

@inproceedings{socher2013recursive,
  title={Recursive deep models for semantic compositionality over a sentiment treebank},
  author={Socher, Richard and Perelygin, Alex and Wu, Jean and Chuang, Jason and Manning, Christopher D and Ng, Andrew Y and Potts, Christopher},
  booktitle={Proc. of the 2013 Conf. on Empirical Methods in Natural Language Processing},
  pages={1631--1642},
  year={2013}
}

@article{sonkiya2021stock,
  title={Stock price prediction using BERT and GAN},
  author={Sonkiya, Priyank and Bajpai, Vikas and Bansal, Anukriti},
  journal={arXiv:2107.09055},
  year={2021}
}

@article{chen2021stock,
  title={Stock movement prediction with financial news using contextualized embedding from bert},
  author={Chen, Qinkai},
  journal={arXiv:2107.08721},
  year={2021}
}

@inproceedings{devlin2018bert,
  author       = {Devlin, Jacob and Chang, Ming-Wei and Lee, Kenton and Toutanova, Kristina},
  title        = {{BERT:} Pre-training of Deep Bidirectional Transformers for Language
                  Understanding},
  booktitle    = {Proc. of the 2019 Conf. of the North American Chapter of
                  the Association for Computational Linguistics: Human Language Technologies},
  pages        = {4171--4186},
  year         = {2019}
}

@article{mantegna1999hierarchical,
  title={Hierarchical structure in financial markets},
  author={Mantegna, Rosario N},
  journal={The European Physical Journal B-Condensed Matter and Complex Systems},
  volume={11},
  pages={193--197},
  year={1999}
}

@inproceedings{ang2022learning,
  title={Learning Dynamic Multimodal Implicit and Explicit Networks for Multiple Financial Tasks},
  author={Ang, Gary and Lim, Ee-Peng},
  booktitle={2022 IEEE Intl. Conf. on Big Data},
  pages={825--834},
  year={2022}
}

@misc{kazemi2019time2vec,
      title={Time2Vec: Learning a Vector Representation of Time}, 
      author={Seyed Mehran Kazemi and Rishab Goel and Sepehr Eghbali and Janahan Ramanan and Jaspreet Sahota and Sanjay Thakur and Stella Wu and Cathal Smyth and Pascal Poupart and Marcus Brubaker},
      year={2019},
      eprint={1907.05321},
      archivePrefix={arXiv},
      
}

@article{boersma2001speak,
  title={Speak and unSpeak with PRAAT},
  author={Boersma, Paul and Van Heuven, Vincent},
  journal={Glot Intl.},
  volume={5},
  pages={341--347},
  year={2001}
}

@inproceedings{gupta2020sentiment,
  title={Sentiment analysis for stock price prediction},
  author={Gupta, Rubi and Chen, Min},
  booktitle={2020 IEEE Conf. on Multimedia Information Processing and Retrieval},
  pages={213--218},
  year={2020}
}

@inproceedings{liu2019combining,
  title={Combining enterprise knowledge graph and news sentiment analysis for stock price prediction},
  author={Liu, Jue and Lu, Zhuocheng and Du, Wei},
  year={2019},
  booktitle = {52nd Hawaii Intl. Conf. on System Sciences},
  pages = {1-9}
}

@article{wang2018combining,
  title={Combining the wisdom of crowds and technical analysis for financial market prediction using deep random subspace ensembles},
  author={Wang, Qili and Xu, Wei and Zheng, Han},
  journal={Neurocomputing},
  volume={299},
  pages={51--61},
  year={2018},
  publisher={Elsevier}
}

@article{schumaker2012evaluating,
  title={Evaluating sentiment in financial news articles},
  author={Schumaker, Robert P and Zhang, Yulei and Huang, Chun-Neng and Chen, Hsinchun},
  journal={Decision Support Systems},
  volume={53},
  number={3},
  pages={458--464},
  year={2012},
  publisher={Elsevier}
}

@article{groth2011intraday,
  title={An intraday market risk management approach based on textual analysis},
  author={Groth, Sven S and Muntermann, Jan},
  journal={Decision Support Systems},
  volume={50},
  number={4},
  pages={680--691},
  year={2011}
}

@inproceedings{butler2009financial,
  title={Financial forecasting using character n-gram analysis and readability scores of annual reports},
  author={Butler, Matthew and Ke{\v{s}}elj, Vlado},
  booktitle={Advances in Artificial Intelligence},
  pages={39--51},
  year={2009}
}

@inproceedings{alostad2015directional,
  title={Directional prediction of stock prices using breaking news on twitter},
  author={Alostad, Hana and Davulcu, Hasan},
  booktitle={2015 {IEEE/WIC/ACM} Intl. Conf. on Web Intelligence and Intelligent Agent Technology},
  volume={1},
  pages={523--530},
  year={2015}
}

@inproceedings{medya2022exploratory,
  title={An Exploratory Study of Stock Price Movements from Earnings Calls},
  author={Medya, Sourav and Rasoolinejad, Mohammad and Yang, Yang and Uzzi, Brian},
  booktitle={Companion Proc. of the Web Conf. 2022 {(WWW'22)}},
  pages={20--31},
  year={2022}
}

@inproceedings{sawhney2021stock,
  title={Stock Selection via Spatiotemporal Hypergraph Attention Network: A Learning to Rank Approach},
  author={Sawhney, Ramit and Agarwal, Shivam and Wadhwa, Arnav and Derr, Tyler and Shah, Rajiv Ratn},
  booktitle={Proc. of the AAAI Conf. on Artificial Intelligence},
  volume={35},
  number={1},
  pages={497--504},
  year={2021}
}

@inproceedings{huynh2023efficient,
  title={Efficient integration of multi-order dynamics and internal dynamics in stock movement prediction},
  author={Huynh, Thanh Trung and Nguyen, Minh Hieu and Nguyen, Thanh Tam and Nguyen, Phi Le and Weidlich, Matthias and Nguyen, Quoc Viet Hung and Aberer, Karl},
  booktitle={Proc. of the 16th ACM Intl. Conf. on Web Search and Data Mining},
  pages={850--858},
  year={2023}
}

@inproceedings{xu2019graph,
  title={Graph wavelet neural network},
  author={Xu, Bingbing and Shen, Huawei and Cao, Qi and Qiu, Yunqi and Cheng, Xueqi},
  booktitle    = {7th Intl. Conf. on Learning Representations {(ICLR'19)}},
  year={2019}
}

@article{wang2022hatr,
  author={Wang, Heyuan and Wang, Tengjiao and Li, Shun and Guan, Shijie},
  journal={IEEE Transactions on Knowledge and Data Engineering}, 
  title={HATR-I: Hierarchical Adaptive Temporal Relational Interaction for Stock Trend Prediction}, 
  year={2023},
  volume={35},
  number={7},
  pages={6988-7002}
}

@inproceedings{sonkiya2022stock,
  title={Stock Price Prediction Using Artificial Intelligence: A Survey},
  author={Sonkiya, Priyank and Bajpai, Vikas and Bansal, Anukriti},
  booktitle={2022 IEEE 7th Intl. Conf. for Convergence in Technology},
  pages={1--9},
  year={2022},
}

@inproceedings{zhan2022meta,
  title={Meta-Adaptive Stock Movement Prediction with Two-Stage Representation Learning},
  author={Zhan, Donglin and Dai, Yusheng and Dong, Yiwei and He, Jinghai and Wang, Zhenyi and Anderson, James},
  year={2022},
  booktitle={NeurIPS 2022 Workshop on Distribution Shifts: Connecting Methods and Applications}
}

@inproceedings{li2022ddg,
  title={DDG-DA: Data Distribution Generation for Predictable Concept Drift Adaptation},
  author={Li, Wendi and Yang, Xiao and Liu, Weiqing and Xia, Yingce and Bian, Jiang},
  booktitle={AAAI Conf. on Artificial Intelligence},
  volume={36},
  number={4},
  pages={4092--4100},
  year={2022}
}

@inproceedings{du2021adarnn,
  title={Adarnn: Adaptive learning and forecasting of time series},
  author={Du, Yuntao and Wang, Jindong and Feng, Wenjie and Pan, Sinno and Qin, Tao and Xu, Renjun and Wang, Chongjun},
  booktitle={Proc. of the 30th ACM Intl. Conf. on Information {\&} Knowledge Management {(CIKM'21)}},
  pages={402--411},
  year={2021}
}

\appendix

\section{Summaries of Datasets}\label{app}

We present an overview of the data sources utilized in the works mentioned in this paper, as shown in Table~\ref{dataset}. 

\begin{table*}[h]
\renewcommand{\arraystretch}{1.3}
\caption{Summaries of Source Data Employed for Knowledge-enhanced Stock Price Prediction}\label{dataset}

\centering
\resizebox{\textwidth}{!}{
\begin{tabular}{ccll}
\toprule
\bfseries Source data &   \bfseries Dataset/API/Platform  & \bfseries URL & \bfseries Paper used\\
\hline

Financial news 
& Sina Weibo & \url{http://weibo.com} or \url{http://finance.sina.com.cn/} & ~\cite{chen2018leveraging, hu2018listening,liu2019combining}\\
& Business Insider & \url{https://www.businessinsider.com} & ~\cite{khan2020stock} \\
& FINET & \url{http://www.finet.hk/mainsite/index.htm} & ~\cite{li2018stock}\\
& Yahoo Finance & \url{https://finance.yahoo.com/} & ~\cite{schumaker2012evaluating}\\
& Epoch News & \url{http://www.epochtimes.com/b5/nsc420.htm} & ~\cite{lee2017predict}\\
& Reuters & \url{http://www.reuters.com/} & ~\cite{huang2022news}\\
& Bloomberg & \url{http://www.bloomberg.com/}& ~\cite{chen2021stock, huang2022news}\\
& Seeking Alpha & \url{https://seekingalpha.com/} & ~\cite{sonkiya2021stock}\\
& Hundsun Electronics & \url{https://www.hundsun.com} & ~\cite{ma2023multi}\\
& Reddit & \url{https://www.reddit.com/r/worldnews/?hl} & ~\cite{deng2019knowledge}\\
& Eastmoney & \url{http://www.eastmoney.com/} & ~\cite{hu2018listening}\\
& Kaggle & \url{https://www.kaggle.com/datasets/} & ~\cite{ang2022learning}\\
& Tencent News& \url{http://news.qq.com/} & ~\cite{liu2019combining}\\
& Phoenix News & \url{http://news.ifeng.com/} & ~\cite{liu2019combining}\\

Social media 
& Twitter & \url{https://developer.twitter.com/en/products/twitter-api} & ~\cite{chen2022gated,khan2020stock, kilimci2020efficient}\\
& Xueqiu & \url{https://www.xueqiu.com/}&~\cite{wang2022hatr}\\
& StockTwits & \url{https://stocktwits.com/} & ~\cite{gupta2020sentiment}\\
& Sina Weibo & \url{http://weibo.com} & ~\cite{wang2018combining}\\

Earnings Call Audio & EarningsCast & \url{https://earningscast.com/} & ~\cite{qin2019you, yang2020html}\\
Earnings Call Transcripts & Seeking Alpha & \url{https://seekingalpha.com/} & ~\cite{qin2019you, yang2020html}\\

& Tushare API & \url{https://tushare.pro/} & ~\cite{zhao2022stock} \\

Industry information 
& GICS & \url{https://en.wikipedia.org/wiki/Global_Industry_Classification_Standard} & ~\cite{cardoso2022learning, sawhney2021stock} \\
& NASDAQ Inc. & \url{https://www.nasdaq.com/screening/industries.aspx} & ~\cite{feng2019temporal, ying2020time}\\
& Tushare API & \url{https://tushare.pro/} & ~\cite{Li2022hyper, xu2021hist}\\
& Yahoo Finance & \url{https://finance.yahoo.com/} & ~\cite{huynh2023efficient}\\

Bond-fund holding data 
& eMAXX & \url{https://emaxx.refinitiv.com/} & ~\cite{zhou2022temporal}\\
Fund-stock holding data & Wind Info & \url{https://www.wind.com.cn/} & ~\cite{lu2018herding}\\
Shareholder data & YFinance API & \url{https://pypi.org/project/yfinance/} & ~\cite{bisarya2022stock}\\ 
Main business of stocks & Tushare API & \url{https://tushare.pro/} & ~\cite{xu2021hist}\\
Financial analysts data & IBES & \url{https://www.refinitiv.com/en/financial-data/company-data/} & ~\cite{creamer2009link} \\
Directors data & IRRC & \url{https://www.irrc.org/} & ~\cite{creamer2009link} \\
Executives data & iFinD & \url{http://www.51ifind.com/} & ~\cite{zhao2022stock} \\
Event data
& GDELT Event DB & \url{https://analysis.gdeltproject.org/module-event-timeline.html} & ~\cite{ang2022learning}\\
Open source KBs & Freebase~\cite{bollacker2008freebase}  & \url{ www.freebase.com} & ~\cite{deng2019knowledge}\\
& Wikidata~\cite{vrandevcic2014wikidata} & \url{https://www.wikidata.org/wiki/Wikidata:List_of_properties/all} & ~\cite{deng2019knowledge, feng2019temporal, sawhney2021stock, sawhney2020deep}\\
& GKG & \url{https://www.gdeltproject.org/} & ~\cite{ang2022learning}\\
\bottomrule
\end{tabular}}
\vspace{-3mm}
\end{table*}

\end{document}